\documentclass[12pt]{article}
\usepackage{amssymb,amsmath,amsthm,epsfig,bm}
\numberwithin{equation}{section}
\numberwithin{figure}{section}
\usepackage{epic,graphicx,color}
\usepackage{ifpdf}
\usepackage{mathrsfs}

\usepackage[noend]{algpseudocode}
\usepackage{algorithmicx,algorithm}

\usepackage{multirow}
\usepackage{makecell}
\usepackage{booktabs}
\usepackage{dsfont}
\usepackage{tabularx}
\usepackage{diagbox}
\usepackage{caption,subcaption}

\usepackage{threeparttable}
\usepackage[numbers,sort&compress]{natbib}

\theoremstyle{plain}

\theoremstyle{definition}

\theoremstyle{remark}
\newtheorem{rem}{Remark}[section]

\newcommand{\n}{\noindent}

\usepackage{indentfirst}

\begin{document}
	
	\title{ \large\bf Convolutional neural network based \n reduced order modeling for multiscale problems}
	
	\author{Xuehan Zhang\thanks{School of Mathematical Sciences,  Tongji University, Shanghai 200092, China. ({\tt  xhzhang@tongji.edu.cn}).}
		\and
		Lijian Jiang\thanks{School of Mathematical Sciences,  Tongji University, Shanghai 200092, China. ({\tt  ljjiang@tongji.edu.cn}).}
	}
\maketitle
\begin{center}{\bf Abstract}
\end{center}\smallskip
In this paper, we combine convolutional neural networks (CNNs) with reduced order modeling (ROM) for efficient simulations of multiscale problems. These problems are modeled by partial differential equations with high-dimensional random inputs. The proposed method involves two separate CNNs: Basis CNNs and Coefficient CNNs (Coef CNNs), which correspond to two main parts of ROM. The method is  called CNN-based ROM. The former one learns input-specific basis functions from the snapshots of fine-scale solutions. An activation function, inspired by Galerkin projection, is utilized at the output layer to reconstruct fine-scale solutions from the basis functions. Numerical results show that the basis functions learned by the Basis CNNs resemble data, which help to significantly reduce the number of the basis functions. Moreover, CNN-based ROM is less sensitive to data fluctuation caused by numerical errors than traditional ROM. Since the tests of Basis CNNs still need fine-scale stiffness matrix and load vector, it can not be directly applied to nonlinear problems. The Coef CNNs can be applied to nonlinear problems and   designed to determine the coefficients for linear combination of basis functions. In addition, two applications of CNN-based ROM are presented, including predicting MsFEM basis functions within  oversampling regions and building accurate surrogates for inverse problems.

\smallskip
{\bf keywords: reduced-basis methods; convolutional neural network; random multiscale problems}

\section{Introduction}
\label{sec1:introduction}
Multiscale problems, which encompass phenomenons characterized by significant variations across multiple scales, are widely applied  in fields such as materials science, environmental science, biomechanics and many others. Examples include the behavior of composite materials, fluid flow in porous media, biological processes and turbulent fluid dynamics. Many of the problems are often  modeled by partial differential equations (PDEs) with random inputs. The randomness is caused by heterogeneity of physical properties. For groundwater, the dispersed phases (e.g., pores or fractures), which may be randomly distributed in the spatial space, lead to fluctuation of hydraulic conductivity \cite{sec1:Darcy_flow_porous}. Moreover, the conductivity difference between phases is typically high-contrast. To capture the fine-scale features of systems, the direct numerical simulations using finite element methods are not feasible due to large scale of computation. Therefore, it is necessary to design efficient multiscale simulation methods, especially for real-time applications (e.g., inverse problems and stochastic control \cite{sec1:inverse problems,sec1:optimal control}) and many-query contexts (e.g., design or optimization \cite{sec1:multiscale design}).

During last decades, the idea of  reduced-order modeling has been proposed for multiscale problems. The construction of reduced-order basis functions is the core of ROM. By fully exploring the local properties of differential operators, a class of multiscale reduction methods use a set of multiscale basis functions, which can effectively capture some fine-scale effects. These multiscale methods include multiscale finite element method (MsFEM) \cite{sec1:MsFEM1,sec1:MsFEM2}, generalized multiscale finite element method (GMsFEM) \cite{sec1:GMsFEM1,sec1:GMsFEM2}, constraint energy minimizing generalized multiscale finite element method (CEM-GMsFEM) \cite{sec1:CEM-GMsFEM1,sec1:CEM-GMsFEM2} and so on. Based on finite element methods and homogenization theory, the convergence of these methods can be rigorously analyzed. In addition, the same basis functions can be used for different source terms  since the construction of multiscale basis functions only involves the differential operators. However, many local problems should be solved for each sample of random inputs to obtain the basis functions, which prevents efficient online computation. Unlike multiscale reduction methods, the reduced-basis methods (RB methods) obtain a universal collection of basis functions by data-driven methods such as greedy algorithm \cite{sec1:greedy-based ROM} and proper orthogonal decomposition (POD) \cite{sec1:POD-based ROM}. However, most existing works have difficulty scaling to high-dimensional problems, such as random inputs based on Gaussian processes \cite{sec1:Gaussian Field}.

To obtain the approximated solutions, ROM usually implements Galerkin projection on a low-dimensional space spanned by the constructed basis functions. Furthermore, techniques, such as Discrete Empirical Interpolation Methods (DEIM) \cite{sec1:DEIM,sec1:MDEIM}, are used to obtain affine decompositions for nonlinear problems and achieve fast online computation. Although DEIM can significantly reduce online computation costs, there are still a couple of limitations. Firstly, a large number of basis functions are required for high-dimensional and highly nonlinear problems. This will impact on the efficiency of DEIM. Secondly, ROM with DEIM is still nonlinear, and need to recall nonlinear solvers such as Newton methods \cite{sec1:nonlinear ROM}.

The ROM for multiscale problems has two stages: the construction of reduced-order basis functions and the computation of coarsen models. To improve the online efficiency, it is desirable  to learn a quick response from samples of random inputs to basis functions. The ROM for multiscale problems assisted by deep learning techniques, which have powerful representation and generalization abilities, gains more and more  attentions. For multiscale reduction methods, M. Wang et al. \cite{sec1:predictionOFdiscretization_GMsFEM} learn the mapping from permeability to GMsFEM basis functions and coarse-scale stiffness matrices on local coarse grids using fully connected neural networks (FNNs). A. Choubineh et al. \cite{sec1:mixed_GMsFEM} reconstruct the basis functions of mixed GMsFEM from permeability matrix using Convolutional Neural Networks (CNNs). For reduced-basis methods, S. Cheung et al. \cite{sec1:DeepGlobalModelReduction} firstly use proper orthogonal decomposition (POD) to construct a set of global nodal basis functions. Then an FNN is used to approximate evolution of the coefficients and, therefore, the porous media flow. In addition, learning the mapping from parameters to coefficients using deep learning techniques is not a new idea in the field of non-intrusive reduced-order modeling \cite{sec1:ROMs1,sec1:ROMs2,sec1:ROMs3,sec1:ROMs4,sec1:ROMs5}.

In this paper, we consider the case that the multiscale problems are represented by steady partial differential equations with random inputs. Our goal is to develop a data-driven method to model multiscale problems and make fast online predictions. Similar to RB methods, we assume that there are snapshots of fine-scale solutions. Two distinct CNNs are designed through following two main stages  of ROM, thus introducing the proposed method as CNN-based ROM. The first ones are  called Basis CNNs, which learn the mapping from samples of random inputs to basis functions. Activation functions, inspired by Galerkin projection methods, are utilized at the output layers to reconstruct fine-scale solutions from the basis functions. Besides, condition number under Frobenius norm are included in loss functions to ensure the stability of training. Note that Basis CNNs can provide input-specific basis functions like multiscale reduction methods, which can break the limitation of RB methods for high-dimensional problems. However, the predictions of Basis CNNs are still dependent on fine-scale stiffness matrix and load vector of FEM, which are not available for nonlinear equations at the online stage. To overcome these issues, we design the second CNN, called Coefficient CNNs (Coef CNNs), to learn the final linear combinations of the approximated solutions using the same dataset as the Basis CNNs.

This article is structured as follows. In Section \ref{sec2:problem setup}, we give the definitions and examples of multiscale problems, which is followed by a brief introduction to the reduced basis methods and convolutional neural networks in Section \ref{sec3:Preliminary}. Section \ref{sec4:Proposed methods} focuses on the proposed method, CNN-based ROM. In this section, we also explore the connections between the proposed method with MsFEM. In Section \ref{sec5:numerical results}, a few numerical results are presented to illustrate the efficacy of the proposed method and compare it with POD-based RB methods. Two applications of CNN-based ROM are also available in this section, including learning MsFEM basis functions with oversampling techniques and being used as surrogates in  inverse problems. Finally, some conclusions are given.
\section{Problem setup}
\label{sec2:problem setup}
In this paper, we consider the following  multiscale problems defined in a bounded domain $\mathcal{S}\subseteq\mathbb{R}^2$ with boundary $\partial{\mathcal{S}}$:
\begin{equation}
	\left\{
	\begin{aligned}
		&\mathcal{L}\Big(u(x);\kappa(x,\xi)\Big)=f(u,x), x \in \mathcal{S},\\
		&\mathcal{D}u(x)=0, x \in \partial{\mathcal{S}},
	\end{aligned}
	\right.
	\label{sec2_eq:problem setup}
\end{equation}
where $\mathcal{L}$ denotes a nonlinear differential operator encoded with random input $\kappa(x,\xi)$. More specifically, $\kappa(x,\xi)$ is a random variable defined on a probability space $\mathcal{P}=(\Omega,\mathcal{F},\mathbb{P})$ for any fixed point $x\in \mathcal{S}$. Besides, $f(u,x)$ is the nonlinear source term with sufficient regularity. The equations (\ref{sec2_eq:problem setup}) thus have unique solutions with the boundary conditions $\mathcal{D}u(x)=0$. Next, we give an example for (\ref{sec2_eq:problem setup}).
\begin{itemize}
	\item \textbf{Diffusion equations}. In this case, we introduce a linear elliptic equation with Dirichlet boundary conditions as follows, 
	\begin{equation}
		\left\{
		\begin{aligned}
			&-\nabla \cdot \big( \kappa(x,\xi)\nabla u(x)\big) =f(x),\ x \in \mathcal{S},\\
			&u(x)=g(x),\ x \in \partial{\mathcal{S}},
		\end{aligned}
		\right.
		\label{sec2_eq:diffusion}
	\end{equation}
	where $\kappa(x,\xi)$ is a permeability field with multiscale features, and $g(x)$ is Dirichlet boundary condition.
\end{itemize}

A numerical approximation of the solution $u$ of the problem (\ref{sec2_eq:problem setup})  can be obtained by finite element methods. Denote by $V$ the appropriate solution space, the variational form of the problem (\ref{sec2_eq:problem setup}) with homogenous Dirichlet boundary reads: find the approximated solution $\hat{u}\in V$ such that
\begin{equation}
	\big( \mathcal{L}(\hat{u}(x);\kappa(x;\xi)),v\big) =\big( f(\hat{u},x),v\big) , \ \forall v\in V,
	\nonumber
\end{equation}
where
\begin{equation}
	\big(\mathcal{L}(\hat{u}(x);\kappa(x;\xi)),v\big) =\int_{\mathcal{S}} \mathcal{L}(\hat{u}(x);\kappa(x,\xi))v(x)dx , \big( f(\hat{u}),v\big) =\int_{\mathcal{S}} f(\hat{u}(x),x) v(x)dx.
	\nonumber
\end{equation}

If we partition the spatial domain by the grid size $h$ (the corresponding mesh grid is denoted as $\mathcal{M}$), we can obtain a finite-dimensional approximation $u_{h}\in V_h \subseteq \mathbb{R}^{N_h\times N_h}$ of $u$ by solving the following nonlinear algebra equation,
\begin{equation}
	A_h(u_h;K)u_h(K)=F_h(u_h;K),
	\label{sec2_eq:variational form2}
\end{equation}
The random input $\kappa$ is discretized over the mesh grid $\mathcal{M}$, thus is equivalent to a high-dimensional random matrix $K\in \mathbb{R}^{N_h\times N_h}$. The notation $K_i$ represents the $i$-th sample of $K$.

To obtain   a  high-fidelity approximation of the problems (\ref{sec2_eq:problem setup}) that have multiscale features, the number of degrees of freedom $N_h$ is usually required to be large. Given the dataset $\{K_i, u_h(K_i)\}_{i=1}^M$, our goal is to find a low-dimensional approximation $u_N$ of $u_h$ to solve the variational problems (\ref{sec2_eq:variational form2}) efficiently, yet retaining the essential features of the maps $K\rightarrow u_h$.

\section{Preliminary knowledge}
\label{sec3:Preliminary}
In this section, we will introduce the preliminary knowledge for the proposed method.  The problem  outlined in Section \ref{sec2:problem setup} will be considered  in  ROM. In Subsection \ref{sec3:RB method}, we give a short review of RB methods, accompanied by an analysis of their limitations. Following this, Subsection \ref{sec3:CNN} will provide a brief introduction to CNNs and compare them with fully connected neural networks.

\subsection{The reduced basis method}
\label{sec3:RB method}
The problem (\ref{sec2_eq:variational form2}) is  a parametrized PDE with high-dimensional parameter space. In the field of model reduction of parametrized PDEs, the RB method has been one of most widely used method. In the following, we briefly review the main ideas of the RB method, which is the foundation of our work. Additional details are referred to \cite{sec3:RB methods1,sec3:RB methods2}.

 The RB method assume that the high-fidelity approximation $u_h$ has low-dimensional features, i.e.,  $u_h$ can be well represented by the linear combination of $N$ basis functions, $N \ll N_h$. The $N$ basis functions construct a space called reduced space $V_N$ (the RB space), which is algebraically denoted by the matrix $P_N\in\mathbb{R}^{N_h\times N}:=[p_1,p_2,\ldots,p_N], \ \ p_i\in\mathbb{R}^{N_h\times 1}, \ \ i=1,\ldots,N$. Then we can obtain the following RB problem
 \begin{equation}
	A_N(u_N;K)u_N(K)=F_N(u_N;K),
	\label{sec3_eq:RB problem}
\end{equation}
where $A_N\in\mathbb{R}^{N\times N}$, $F_N\in\mathbb{R}^{N\times 1}$ and $u_N(K)\in\mathbb{R}^{N\times 1}$ is the reduced vector of degrees of freedom that is called RB solution. A common criterion to obtain the RB problem (\ref{sec3_eq:RB problem}) from (\ref{sec2_eq:variational form2}) is Galerkin projection \cite{sec3:Galerkin projection}, that is
\begin{equation}
	A_N(u_N;K)= P_N^TA_h(P_N^Tu_N;K)P_N, \ F_N(u_N;K)= P_N^TF_h(P_N^Tu_N;K).
	\label{sec3_eq:Garlerkin}
\end{equation}

The computational procedure of RB methods is commonly decomposed into offline and online phases: during offline phase,  reduced basis functions are generated, and auxiliary quantities are precomputed. Then, in the online phase, for different instances of the parameters, the RB solutions can be efficiently computed.

\subsubsection{The construction of RB space}
\label{ssec3:The construction of RB space}
The goal of RB methods is to find linear approximation spaces $V_N$ for which the worst best-approximation error for elements of $V_h$,
\begin{equation}
	d_{V_N}(V_h):=\mathop{\sup}\limits_{u_h\in V_h}\mathop{\inf}\limits_{\nu\in V_N}\Vert u_h-\nu\Vert,
	\nonumber
\end{equation}
is near the Kolmogorov N-width of $V_h$
\begin{equation}
	d_{N}(V_h):=\mathop{\inf}\limits_{\substack{W\subseteq V_h \\ \dim W\le N}}\mathop{\sup}\limits_{u_h\in V_h}\mathop{\inf}\limits_{\nu\in W}\Vert u_h-\nu\Vert.
	\nonumber
\end{equation}
The existence of the optimal subspace $V_N$ is guaranteed by the Theorem II.2.3 in \cite{sec3:RB_existence}.

In practical, the reduced basis functions $P_N$ of $V_N$ can be generated by either greedy algorithms or proper orthogonal decomposition (POD). We briefly recall the latter as we will use it in the numerical results for comparision. Given a snapshot matrix of data
\[
\mathcal{U} = \{u_h(K_1),u_h(K_2),\ldots,u_h(K_M)\}
 \]
 the main idea of POD is to find a finite subspace $V_N\subseteq V_h$ with orthogonal basis $P_N:=[p_1,p_2,\ldots,p_N]$ to minimize the projection error,
\begin{equation}
	P_N = \mathop{\arg\min}\limits_{\substack{P^TP=I \\ p_i\in V_N}}\sum_{i=1}^M \Vert u_h(K_i) - PP^Tu_h(K_i) \Vert_2^2.
	\nonumber
\end{equation}
By taking advantage of the singular value decomposition (SVD) of $\mathcal{U}$
\begin{equation}
	U = P\Sigma Z^T,
	\nonumber
\end{equation}
we can obtain the matrix $P_N$ whose columns are the $N$ orthogonal basis of $V_N$: that is the $N$ columns of $P$ corresponding to the largest diagonal elements in $\Sigma$.

We note that POD is a powerful tool to exploit the low-rank features of data. However, there are still two bottlenecks:
\begin{itemize}
	\item POD only provide the optimal linear representations of data in $l_2$ sense, and it is not guaranteed that they are the best basis functions in a broader sense.
	\item RB methods, including POD-based RB methods, try to represent the solutions for different parameters in the same subspace of $V_h$, which means that a larger number of basis functions are needed to obtain accurate approximation of parametrized PDEs with high-dimensional parameter space. This will affect the efficiency and accuracy of RB methods.
\end{itemize}

\subsubsection{Online stage of RB methods}
Efficient implementation of RB-methods relies on the affine decompositions of stiffness matrix and load vectors, that is $A_h(u_h;K)$ and $F_h(u_h;K)$ can be written in parameter-seperable form, as follows
\begin{equation}
A_h(P_N^Tu_N;K)=\sum_{q=1}^{Q_A}c_A^q(u_N,K)A_h^q, \ F_h(P_N^Tu_N;K)=\sum_{q=1}^{Q_F}c_F^q(u_N,K)F_h^q.
\nonumber
\end{equation}
According to (\ref{sec3_eq:Garlerkin}), we can obtain that
\begin{equation}
	A_N(u_N;K)=\sum_{q=1}^{Q_A}c_A^q(u_N,K)A_N^q, \ F_N(u_N;K)=\sum_{q=1}^{Q_F}c_F^q(u_N,K)F_N^q,
	\nonumber
\end{equation}
where $A_N^q =P_N^T A_h^q P_N, F_N^q =P_N^T F_h^q$. Precomputing $\{A_N^q\}_{q=1}^{Q_A}$ and $ \{F_N^q\}_{q=1}^{Q_F}$ in the offline phase, we can avoid operations with a complexity dependent on $N_h$.

However, affine decompositions of bilinear forms (or linear forms) is a strong assumption, which is not satisfied in many cases. DEIM are used to provide an approximated parameter-seperable representations. During offline phase, POD is used to obtain parameter-independent functions $\{A_N^q\}_{q=1}^{Q_A}$ and $ \{F_N^q\}_{q=1}^{Q_F}$. Then, during online phase, an interpolation problem is solved to obtain the coefficients $\{c_N^q(u_N,K)\}_{q=1}^{Q_A}$ and $ \{c_N^q(u_N,K)\}_{q=1}^{Q_F}$ for each new parameters $K$.

DEIM techniques are powerful tool to deal with nonlinear and non-affine RB problems, however there are two main limitations:
\begin{itemize}
	\item For complex problems (i.e., high-dimensional and nonlinear problems), a large number of basis functions must be employed to obtain an accurate approximation, which prevents efficiency of online computation.
	\item For nonlinear problems, DEIM can help to reduce the computation costs but can not avoid iteration (such as newton iteration methods) since it is an interpolation method.
\end{itemize}

\subsection{Convolutional Neural Network}
\label{sec3:CNN}
\begin{figure}[ht]
	\centering
	\subcaptionbox{Fully connections.\label{sec3_fig:FNN}}
	{\includegraphics[scale=0.15]{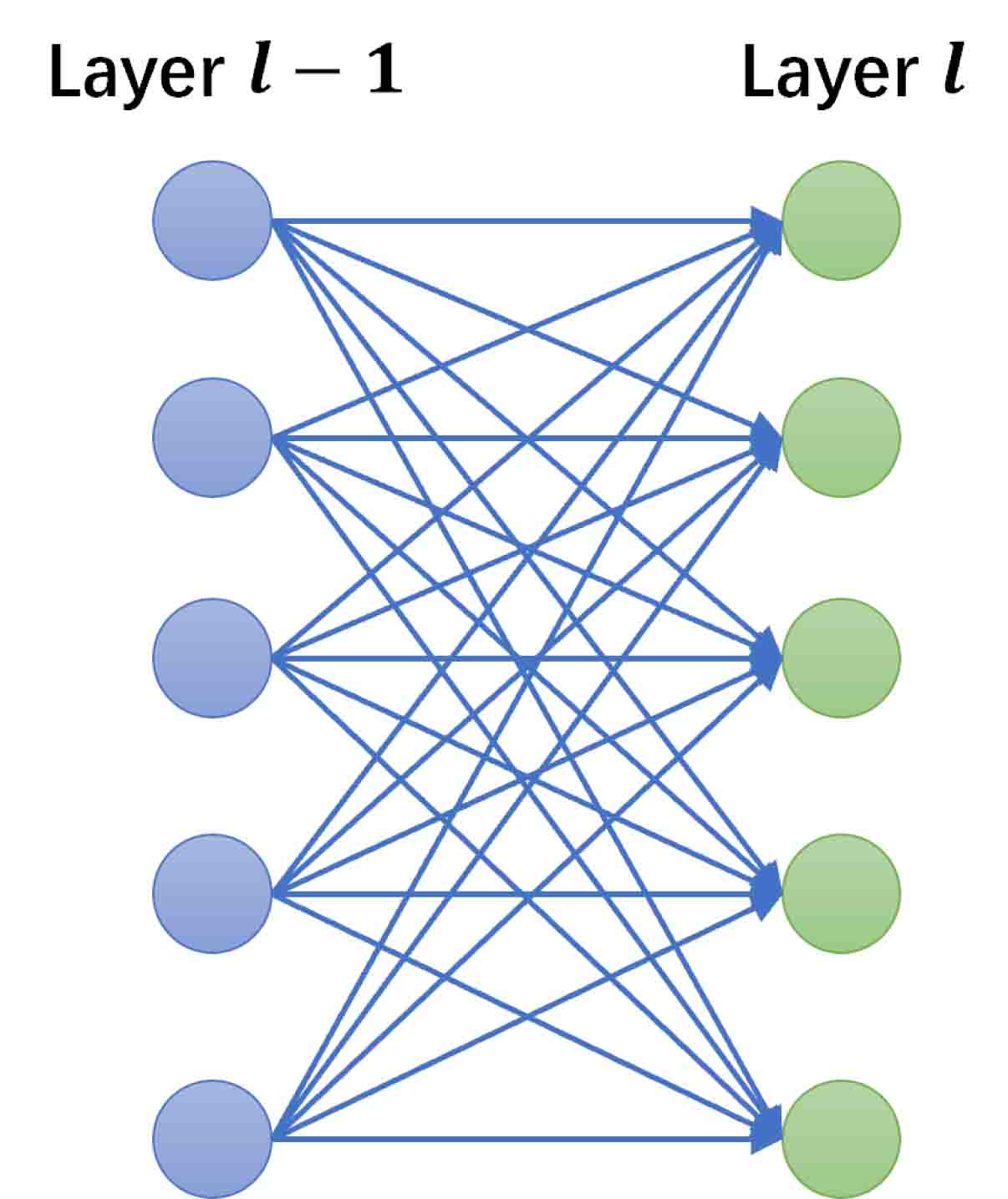}}
	\quad\quad
	\subcaptionbox{Local connections.\label{sec3_fig:CNN_local_connection}}
	{\includegraphics[scale=0.15]{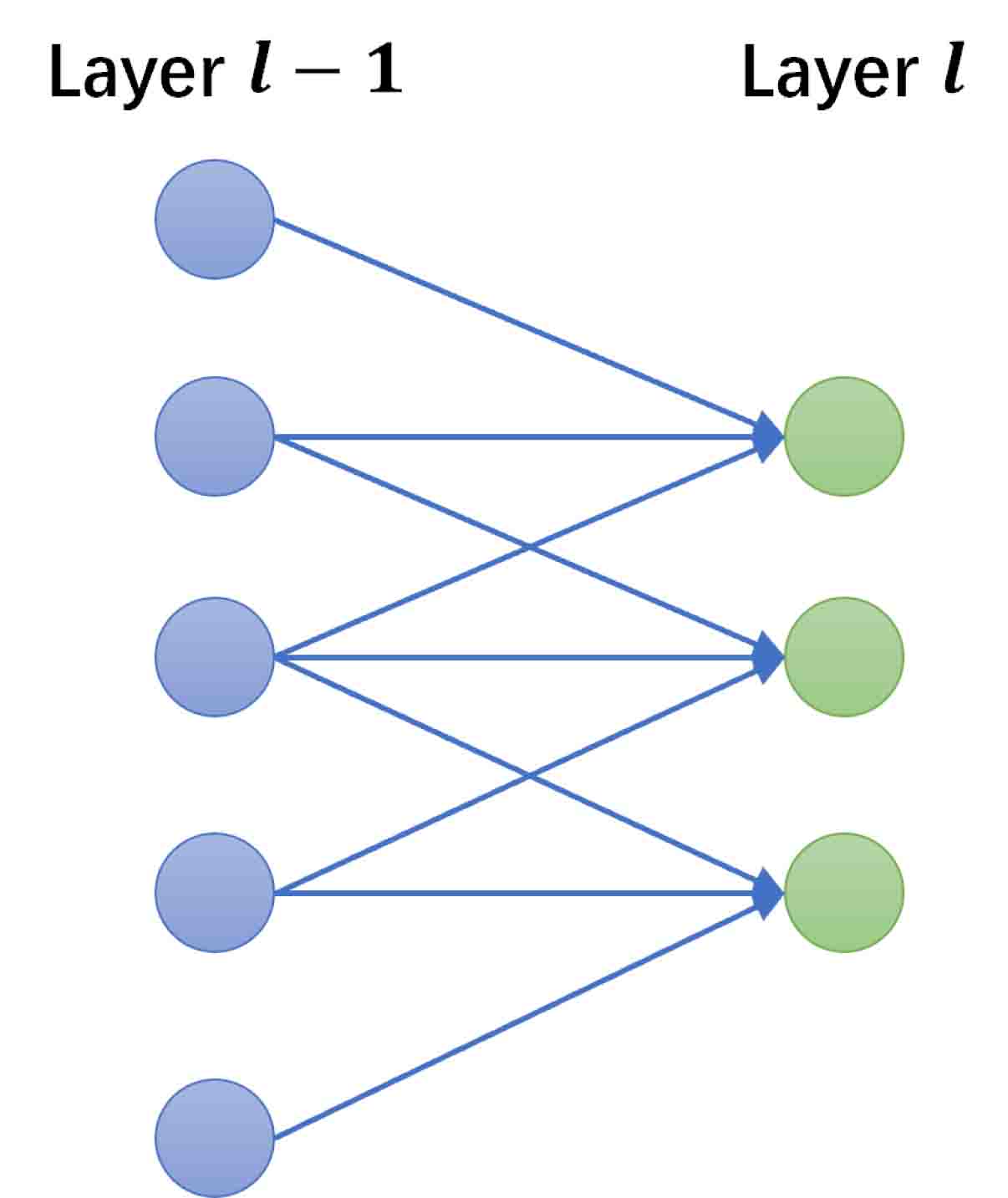}}
	\quad\quad
	\subcaptionbox{Shared weights.\label{sec3_fig:CNN_shared_weights}}
	{\includegraphics[scale=0.15]{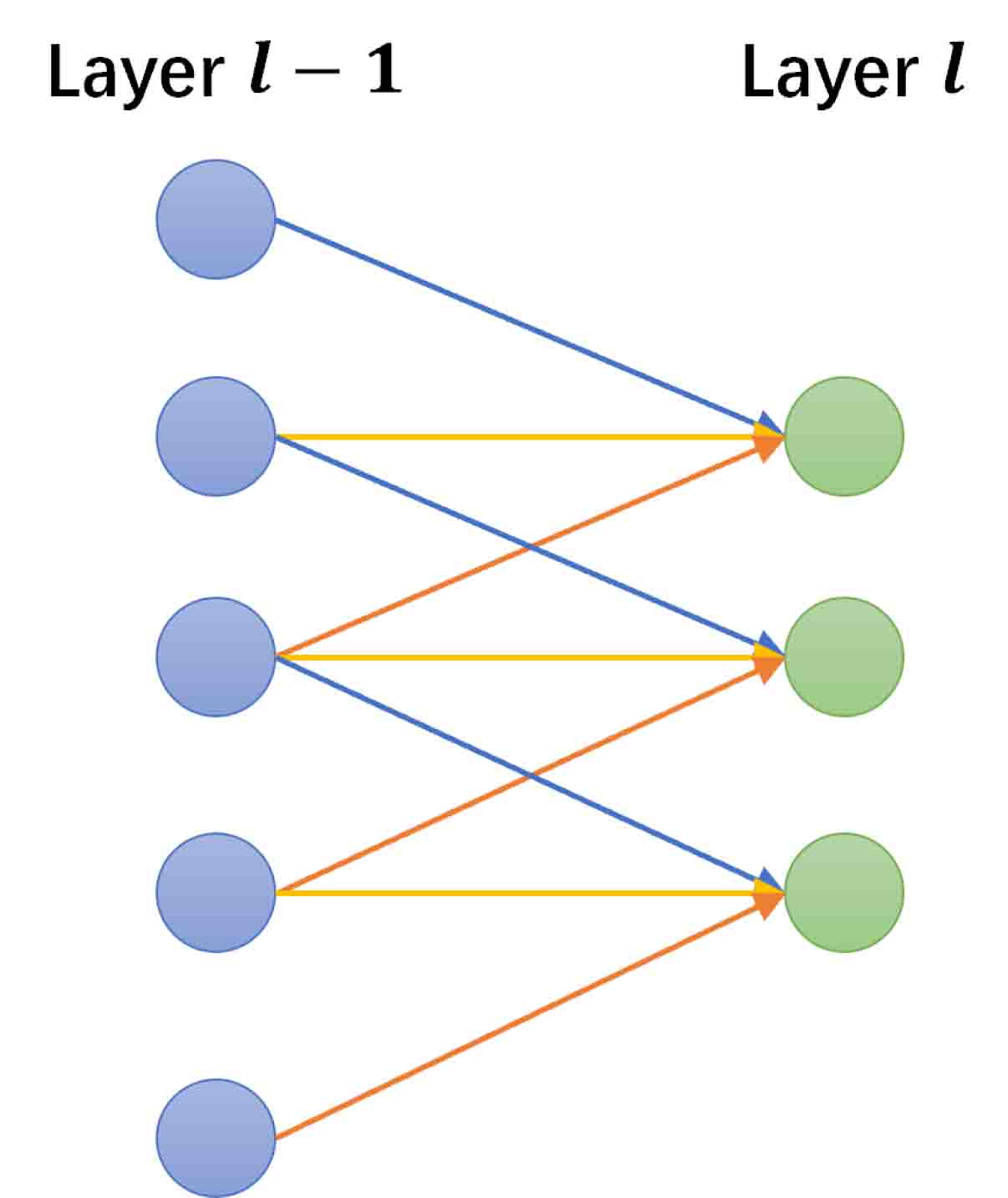}}
	\caption{Architectures of the FNN and the CNN.}
\end{figure}
In the field of machine learning, the problem (\ref{sec2_eq:problem setup}) can be regarded as an image-to-image regression task \cite{sec1:Gaussian Field}. Both the discrete solution $u_h$ and the sample of the random input $K$ are matrix with respect to FE mesh $\mathcal{M}$, which are usually high-dimensional. Using feed-forward neural networks (FNN) for such tasks usually leads to large training parameters and is prone to overfitting, because they are usually fully-connected, that is, each neuron in one layer is connected to all neurons in the next layer (see Figure \ref{sec3_fig:FNN}). Convolutional neural networks is a special type of deep neural network that is inspired by human visual system \cite{sec3:CNN1} and have been widely used in many fields such as image recognition, natural language processing and so on. There are two special aspects in the architecture of CNN, i.e., local connections (see Figure \ref{sec3_fig:CNN_local_connection}) and shared weights (see Figure \ref{sec3_fig:CNN_shared_weights}), which can well deal with image-to-image regression tasks.

\begin{figure}[ht]
	\centering
	\includegraphics[scale=0.11]{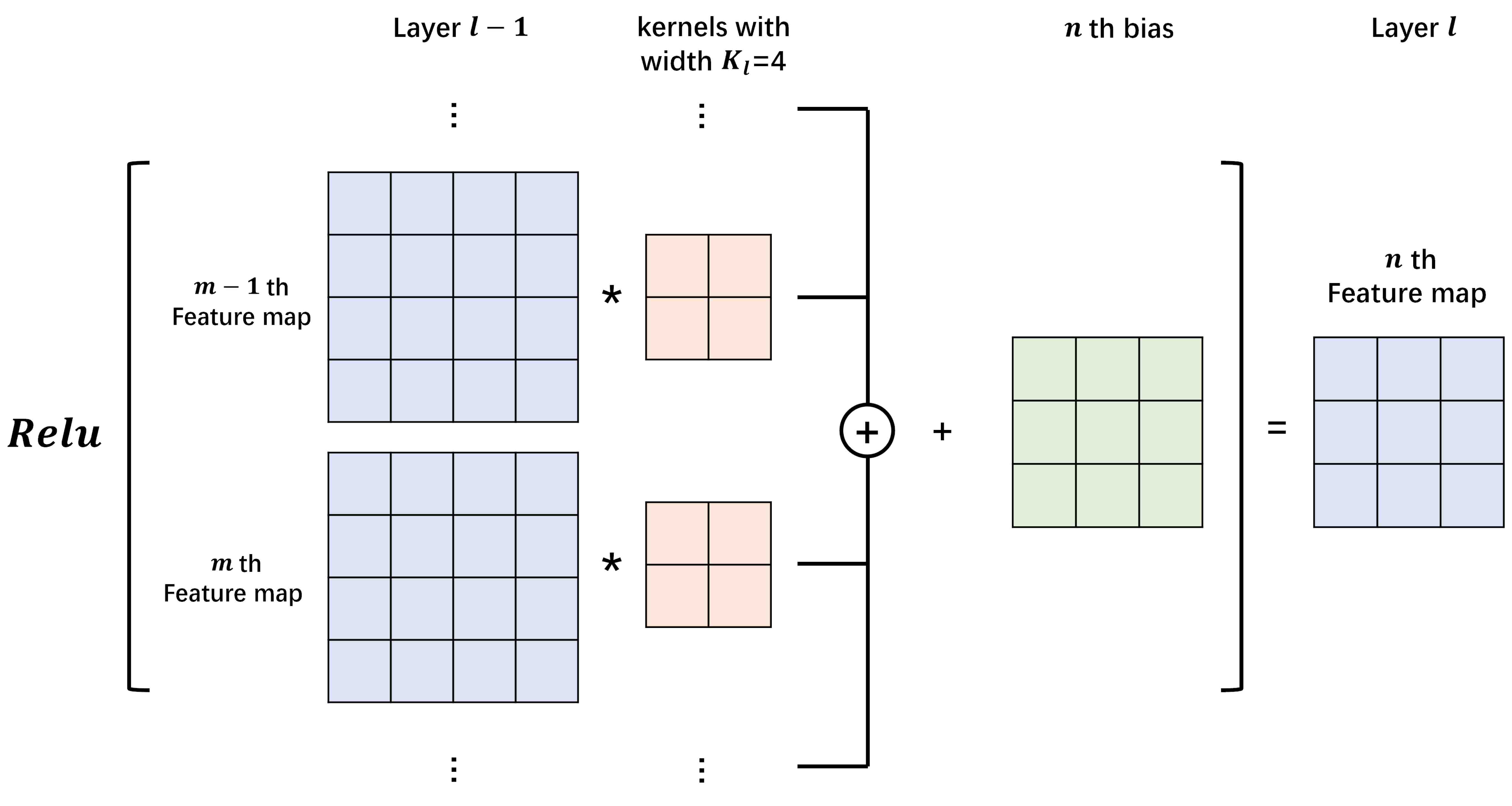}
	\caption{Convolutional layer in the CNN.}
	\label{sec3_fig:convolutional_Layer}
\end{figure}

A complete CNN contains convolution layers and pooling layers. We briefly introduce the former as we will use it in the proposed methods. The value of a neuron $y_{ln}^x$ at position x of the $n\ $th feature map in the $l\ $th layer is defined as (see Figure \ref{sec3_fig:convolutional_Layer})
\begin{equation}
	y_{ln}^x = \sigma\Big(b_{ln}+\sum_{m}\sum_{k=0}^{K_l-1}w_{lnm}^{k}y_{(l-1)m}^{x+k}\Big),
	\nonumber
\end{equation}
where $m$ indexes the feature map in the previous layer ($l-1\ $th layer), $w_{lnm}^{k}$ is the weight of position $x$ connected to the $m\ $th feature map, $K_l$ is the width of the kernel, $b_{ln}$ is the bias of the $n\ $th feature map in current layer, and $\sigma$ is the activation function.

As shown in Figure \ref{sec3_fig:convolutional_Layer}, convolution operations will change the size of inputs. To preserve the spatial dimensions of the inputs or feature maps, padding techniques \cite{sec3:CNN2}, which refers to the addition of extra pixels around the borders of the input images or feature map, can be used.
\section{Convolutional neural network based reduced-order modeling (CNN-based ROM)}
\label{sec4:Proposed methods}
In this section, we will introduce a novel framework for approximating PDEs with random inputs by coupling CNNs with the ideas of the reduced-order modeling, which is called CNN-based ROM. There are two main components of the proposed architecture, that is Basis CNNs and Coefficient CNNs (Coef CNNs). We first learn the input-specific basis functions $P_N(K)$ by Basis CNNs. Besides dataset $\Big\{ K_i, u_h(K_i)\Big\} _{i=1}^M$, FEM bilinear forms (or linear forms) $\Big\{ Ah(uh(K_i);K_i),Fh(uh(K_i);K_i)\Big\} _{i=1}^M$, as parts of activation functions at the output layer, are involved to obtain the reduced-order solution $u_N$. However, such Basis CNNs alone can only be used to linear problems since unknowns $u_h(K_i)$ still exists in the FEM bilinear forms (or linear forms). In terms of the results of Basis CNNs, Coef CNNs are thus designed to solve these issues through learning the mapping from $K$ to coefficients, which can also improve the online efficiency.
\vspace{0.5em}
\begin{rem}
To simplify the notations, we will describe CNN-based ROM for linear problems, i.e.,
\begin{equation}
	A_h(K)u_h(K)=F_h(K).
	\label{sec4_eq:FOM}
\end{equation}
However, the proposed framework includes the nonlinear cases (\ref{sec2_eq:variational form2}), since we can linearize the discrete equations by replacing $u_h$ in $A_h(u_h;K_i)$ and $F_h(u_h;K_i)$ with data $u_h(K_i)$.
\end{rem}
\begin{figure}[ht]
	\centering
	\includegraphics[scale=0.11]{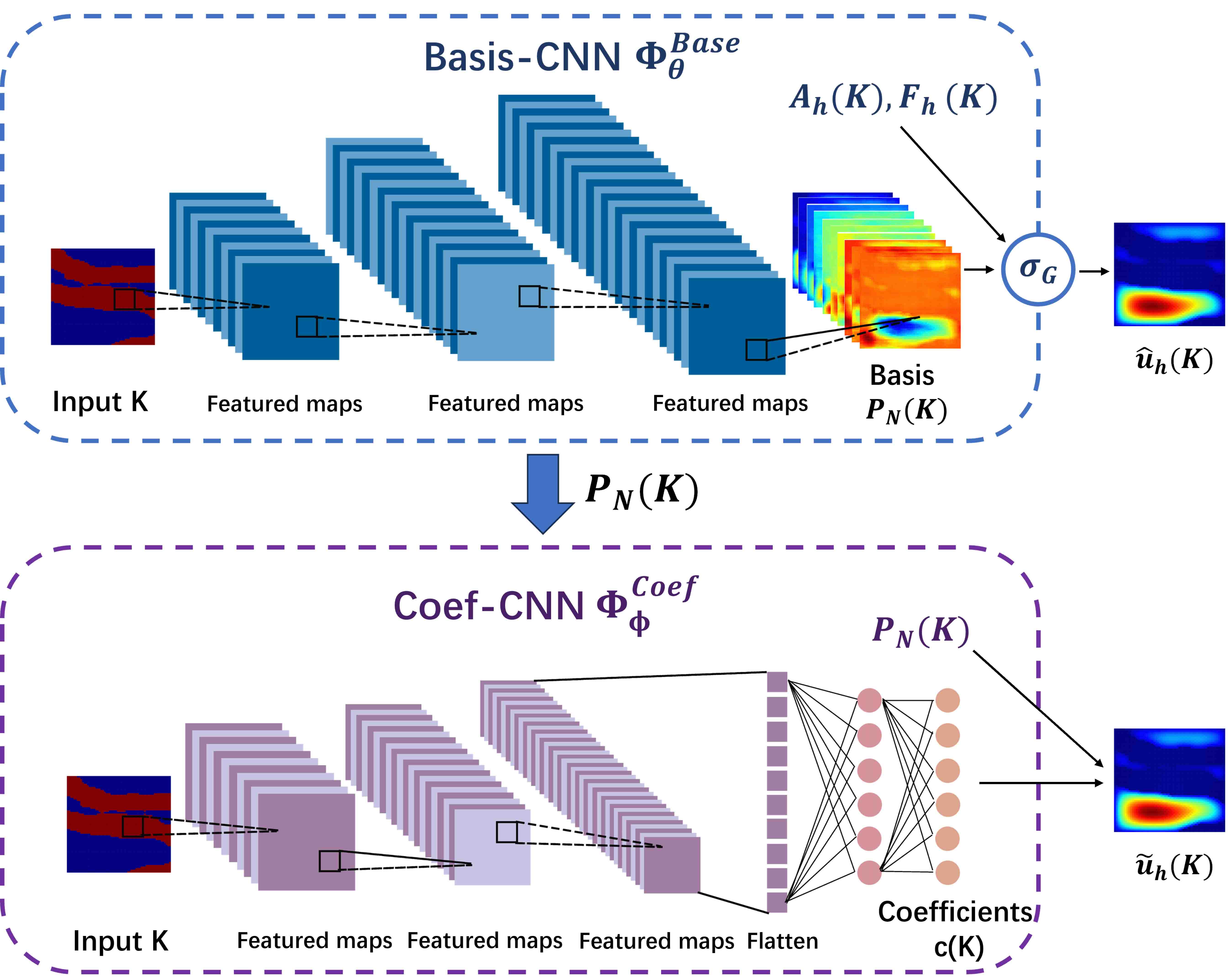}
	\label{sec4_fig:outline}
	\caption{The schema of CNN-based ROM.}
\end{figure}

\subsection{The Basis CNNs (Basis CNNs)}
\label{ssec4:Basis CNN}
PDEs with random inputs theoretically have infinite dimensional parameter space. In practical, to capture the multiscale features in the random inputs, we usually sample from random inputs $\kappa$ at a very fine FEM mesh $\mathcal{M}$. Thus, the problems we considered here have higher dimensional parameter space than ones in traditional RB methods. The success of RB methods significantly depends on the quality of RB basis functions, which can remain the features of the mapping $K\rightarrow u_h$ while realizing model reduction. This requires the RB methods capture the common features of PDEs in the same parameter space, therefore it is crucial to choose representative points in parameters space to form the reduced basis functions, which is intractable in high-dimensional parameter space. A large number of reduced basis functions are often used for high-dimensional and nonlinear problems.

For multiscale reduction methods, a set of local problems require to be solved at the online stage to obtain the basis functions, which may be time-consuming for large-scale problems. In addition, basis functions constructed by multiscale reduction methods also lack global information due to localization.

\begin{figure}[ht]
	\centering
	\subcaptionbox{Convolutional block.\label{sec4_fig:conv_p_block}}
	{\includegraphics[scale=0.17]{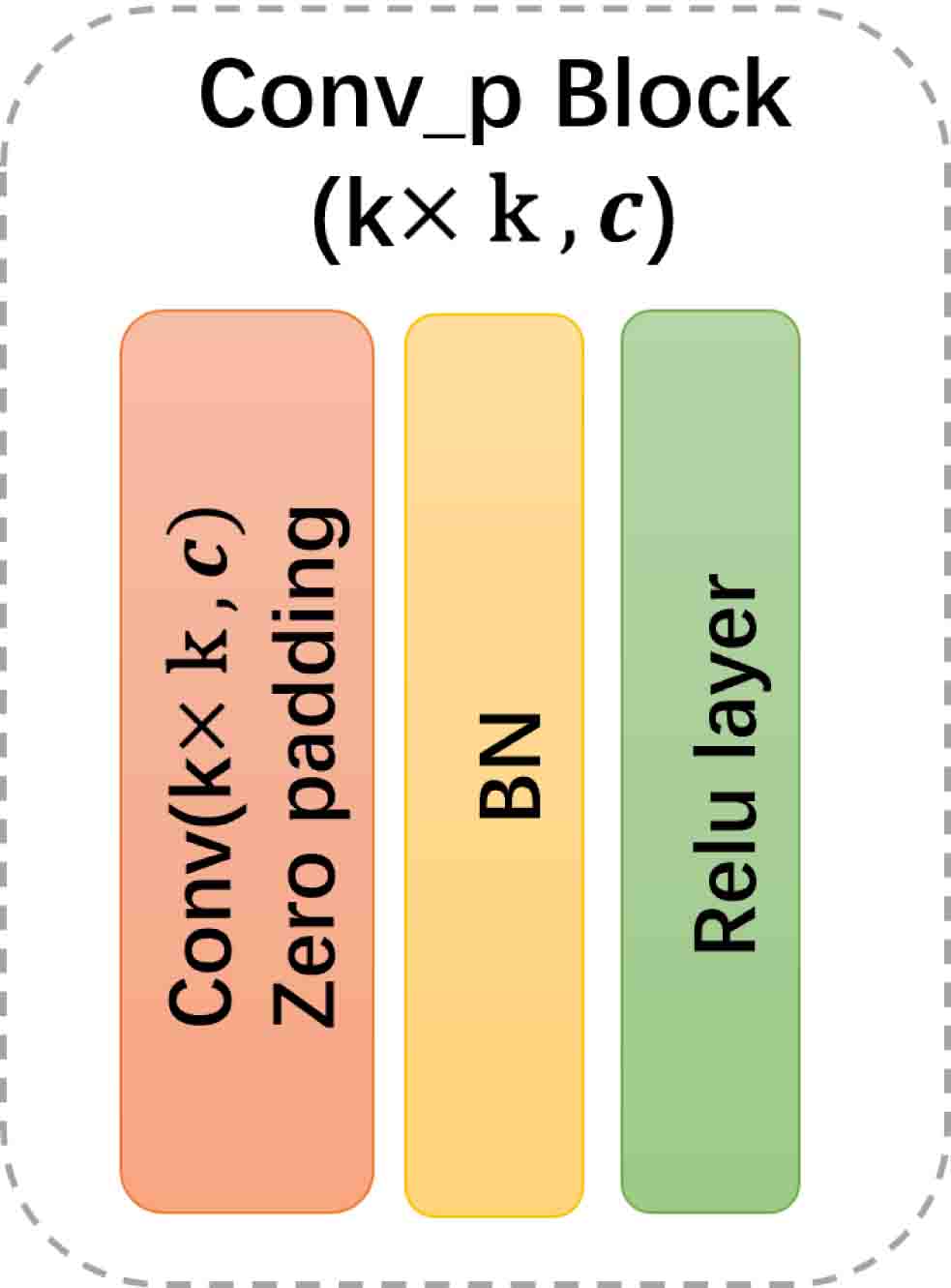}}
	\quad\quad\quad
	\subcaptionbox{Total architecture.\label{sec4_fig:CNN_Base}}
	{\includegraphics[scale=0.14]{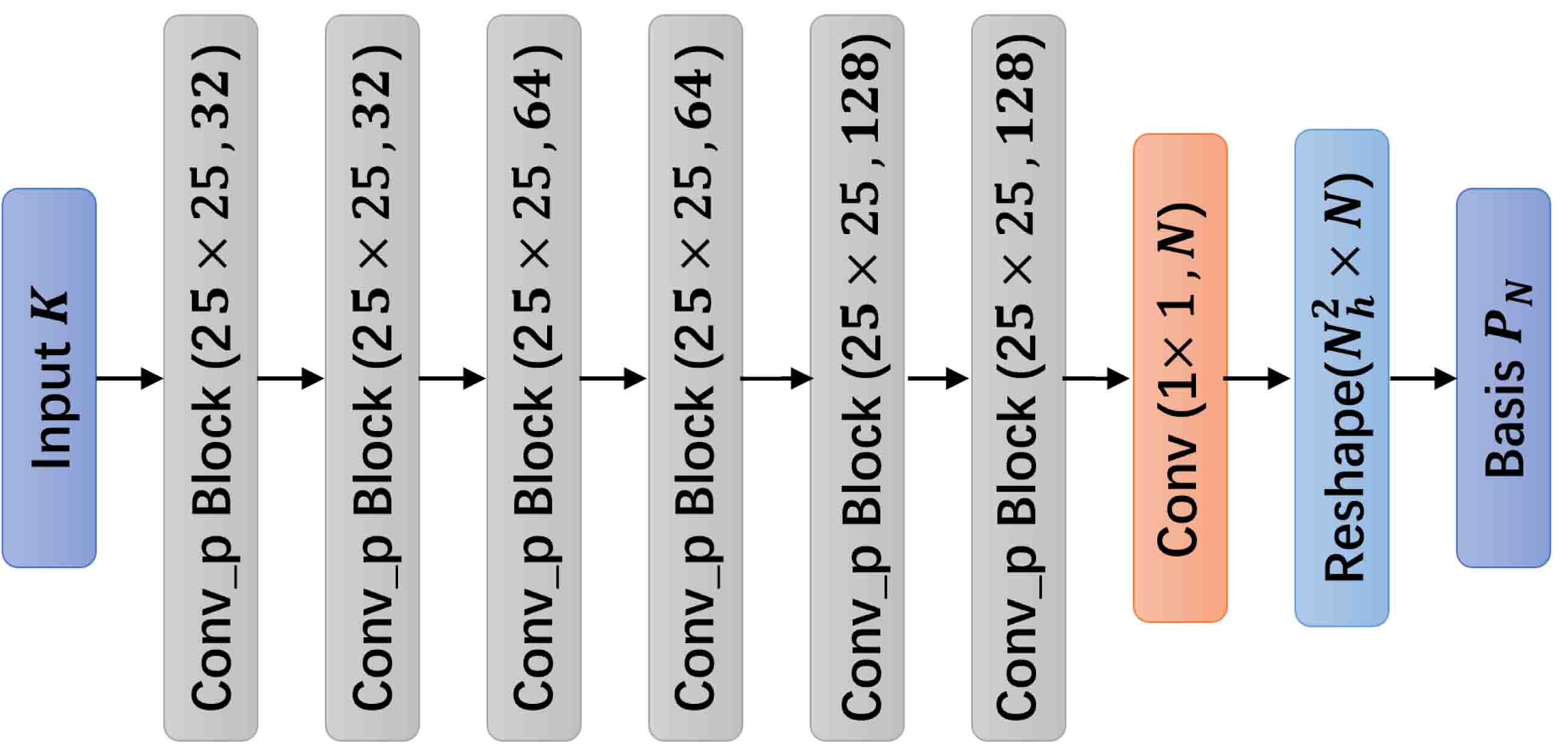}}
	\caption{The architecture of Basis CNNs.}
\end{figure}

To solve these issues, we learn the basis functions from data for each instance of random inputs, i.e.,
\begin{equation}
	\begin{aligned}
		\Phi^{Base}_{\theta}&: \mathbb{R}^{N_h\times N_h}\rightarrow\mathbb{R}^{N_h^2\times N},\\
		P_N(K) &= \Phi^{Base}_{\theta}(K).
	\end{aligned}
	\nonumber
\end{equation}
To capture the multiscale features from image-like inputs $K$, CNN is a good choice. Compared with FNN, local connections and shared weights in the CNN can help better extract significant local features and significantly reduce the trainable parameters. To maintain the integrity of spatial information, pooling layers are not included in Basis CNNs and padding techniques are used to keep the size of feature maps after convolution operations. In addition, we apply Batch Normalization layers (BN) after the convolutional layers (Conv) and before the activation functions (see Figure \ref{sec4_fig:conv_p_block}) to accelerate model convergence and avoid overfitting \cite{sec4:BN}. In the last layer, $1\times1$ convolutional layer are used to adjust the number of basis functions and fuse features of different channels. The specific architecture of Basis CNNs refer to Figure \ref{sec4_fig:CNN_Base}.

\subsubsection{Galerkin-projection activation function}
\label{ssec4:Galerkin-projection activation function}
In this subsection, we will introduce an activation function at the output layer to reconstruct the high-fidelity solution $u_h(K)$ from basis functions, that is, using $\hat{u}_h(K)=P_N(K)u_N(K)$ to approximate the high-fidelity solutions $u_h(K)$. Given $P_N(K)$ by Basis CNNs, we use Galerkin projection to obtain $u_N(K)$ by solving the algebra equation (\ref{sec3_eq:RB problem}). Define the residual by
\begin{equation}
	\mathcal{R}(u_h,K)= A_h(K)u_h(K)-F_h(K).
	\nonumber
\end{equation}
Forcing the residual of $\hat{u}_h$ orthogonal to $V_N$, $u_N(K)$ satisfied
\begin{equation}
	P_N(K)^T\mathcal{R}(\hat{u}_h,K)=P_N(K)^T\Big(A_h(K)P_N(K)u_N(K)-F_h(K)\Big)=0.
	\nonumber
\end{equation}
We can then obtain the approximated solution $\hat{u}_h$ by solving above algebra equation, i.e.,
\begin{equation}
	\hat{u}_h(K)=P_N(K)\Big(P_N(K)^TA_h(K)P_N(K)\Big)^{-1}P_N(K)^TF_h(K),
	\nonumber
\end{equation}
where $P_N(K)=\Phi_{\theta}^{Base}(K)$. We call it Galerkin-projection activation functions $\sigma_{G}$ and define it as a function of $P_N$, that is
\begin{equation}
	\sigma_{G}(P_N;A_h,F_h) =P_N(P_N^TA_hP_N)^{-1}(P_N^TF_h)= P_N A_N^{-1}F_N,
	\nonumber
\end{equation}
where $A_N = P_N^TA_hP_N,\ F_N = P_N^TF_h$.

Above all, We can write the forward propagation of Basis CNNs as
\begin{equation}
	\hat{u}_h(K) = \sigma_{G}\circ \Phi_{\theta}^{Base}(K).
	\label{sec4_eq:forward propagation}
\end{equation}

\begin{rem}
	\label{sec4_rem:cea}
	For elliptic equations, such as the diffusion equations (\ref{sec2_eq:diffusion}), $C\acute{e}a$ lemma \cite{sec4:cea} gaurantees that $\hat{u}_h(K)$ defined in (\ref{sec4_eq:forward propagation}) best approximate the truth $u(K)$ in the linear space $V_N$ spanned by $P_N$. More formally, we denote the bounded and $V_N$-elliptic bilinear form by $a(\cdot,\cdot)$, which leads to the definition of the norm $\Vert u\Vert_{V_N}=\sqrt{a(u,u)}$. Then there is a constant $C>0$, such that
	\begin{equation}
		\Vert u(K)-\hat{u}_h(K)\Vert_{V_N}\le C\mathop{\inf}\limits_{\nu\in V_N}\Vert u(K)-\nu\Vert_{V_N}.
		\nonumber
	\end{equation}
\end{rem}

\begin{rem}
	\label{sec4_rem:derivatives}
	When training the networks, back-propagation algorithms require the derivatives of the loss function with respect to trainable parameters. Therefore, a well-defined activation function $\sigma_{G}$ must be differentiable with respect to input variables, i.e., reduced basis functions $P_N=[p_{ij}]_{N_h^2\times N}$. The $j$-th column of $P_N$ is represented as $p_j, j=1,\ldots,N$. Denote the reduced-order solution by $u_N=A_N^{-1}F_N$ and the elements of matrix $A_h$ by $A_h=[a_{ij}]_{N_h^2\times N_h^2}$, we can compute them as follows
	\begin{equation}
		\begin{aligned}
		\frac{\partial{\sigma_{G}}}{\partial{p_{ij}}} &= u_N*e_j-P_NA_N^{-1}\frac{\partial{A_N}}{\partial{p_{ij}}}A_N^{-1}F_N+P_NA_N^{-1}F_N*e_j,\\
		& = u_N*e_j-P_NA_N^{-1}H_{ij}A_N^{-1}F_N+P_NA_N^{-1}F_N*e_j,
		\end{aligned}
		\nonumber
	\end{equation}
	where $e_i=[0,\ldots,1,\ldots,0]^T\in\mathbb{R}^{N\times 1}$ is the $i$-th unit vector. Besides, the element $h_{kt}$ of matrix $H_{ij}\in \mathbb{R}^{N\times N}$ are
    \begin{equation}
		h_{kt} = \frac{\partial p_k^T A_h p_t}{\partial p_{ij}} = \sum_{m=1}^{N_h^2}\sum_{n=1}^{N_h^2}\frac{\partial p_{mk}p_{nt}a_{mn}}{\partial p_{ij}}.
		\nonumber
	\end{equation}
\end{rem}

\subsubsection{Loss function}
Given the dataset $\Big\{ K_i,u_h(K_i)\Big\} _{i=1}^M$ and FEM bilinear forms (or linear forms) $\Big\{ A_h(K_i),F_h(K_i)\Big\} _{i=1}^M$, we define the loss function as follows
\begin{equation}
	\begin{aligned}
	Loss_{RB}(\theta) &= \frac{1}{M}\sum_{i=1}^M \Vert u_h(K_i)-\hat{u}_h(K_i)\Vert_{L_2}^2+\lambda_{G}\ cond_F\Big(A_N(K_i)\Big)^2,\\
    &= \frac{1}{M}\sum_{i=1}^M \Vert u_h(K_i)-\sigma_{G}\big(\Phi_{\theta}^{Base}(K_i);A_h(K_i),F_h(K_i)\big)\Vert_{L_2}^2\\
	&+\lambda_{G}\ cond_F\Big(\big(\Phi_{\theta}^{Base}(K_i)\big)^TA_h\Phi_{\theta}^{Base}(K_i)\Big)^2,
	\end{aligned}
	\label{sec4_eq:loss function(Base)}
\end{equation}
where the $L_2$ is the $L_2$ norm. In practical, we approximate the $L_2$ norm in the loss function at the discrete points on the FEM mesh $\mathcal{M}$, and $cond_F$ is the condition number under Frobenious norm, i.e.,
\begin{equation}
	cond_F(A) = \Vert A \Vert_F\Vert A \Vert_F^{-1}.
	\nonumber
\end{equation}
The first term of loss function (\ref{sec4_eq:loss function(Base)}) trains the neural networks to approximate the data. As shown in Remark \ref{sec4_rem:derivatives}, the training depends on the inverse of $A_N$. However, $A_N$ learned by neural networks may be ill-conditioned because it does not have good properties, such as orthogonality \cite{sec4:orthonormalization}. To guarantee the stability of training and the reduced-order model (\ref{sec3_eq:RB problem}), we impose the second term in the loss function. In theoretical analysis, condition number under 2-norm are widely used, but it is intractable to compute during training. Therefore, we choose the condition number under Frobenious norm here. The stability parameter $\lambda_{G}$ is a hyperparameter.

\subsection{The Coefficient CNNs (Coef CNNs)}
\label{ssec4:Coef CNN}
The Basis CNNs can generate a reduced-order basis function $P_N$ of the problem (\ref{sec4_eq:FOM}) for given input $K$, and the corresponding reduced-order model can be written as follows
\begin{equation}
	A_N(K)u_N(K)=F_N(K),
	\label{sec4_eq:ROM}
\end{equation}
where $A_N(K)=P_N^TA_h(K)P_N$ and $F_N(K)=P_N^TF_h(K)$. Here, the computation of $A_N(K)$ and $F_N(K)$ still depend on $N_h^2$, which prevents the efficiency of online stage. Besides, we should note that Basis CNNs alone is not applicable for nonlinear equations because $A_h$ and $F_h$ not only depend on input $K$ but also unknown solution $u_h$. To overcome these issues and extend the idea to nonlinear problems, Coef CNNs are designed to learn the coefficients $c(K)$ of linear combination $\tilde{u}_h(K)=\Phi_{\theta}^{Base}(K)c(K)$ to approximate high fidelity solution $u_h(K)$, i.e.,
\begin{equation}
	\begin{aligned}
	\Phi_{\phi}^{Coef}&:\mathbb{R}^{N_h\times N_h}\rightarrow \mathbb{R}^{N \times 1},\\
	c(K) & = \Phi_{\phi}^{Coef}(K),
	\end{aligned}
	\nonumber
\end{equation}
where $c(K)$ is a $N$ -dimensional vector with elements $c_i(K), i=1,\ldots,N$. Combined with the result of Basis CNNs, we can obtain the following surrogates of the equations (\ref{sec2:problem setup}),
\begin{equation}
	\tilde{u}_h(K)=P_N(K)c(K)=\Phi_{\theta}^{RB}(K)\Phi_{\phi}^{Coef}(K).
	\nonumber
\end{equation}

\begin{figure}[ht]
	\centering
	\subcaptionbox{Convolutional block.\label{sec4_fig:conv_block}}
	{\includegraphics[scale=0.18]{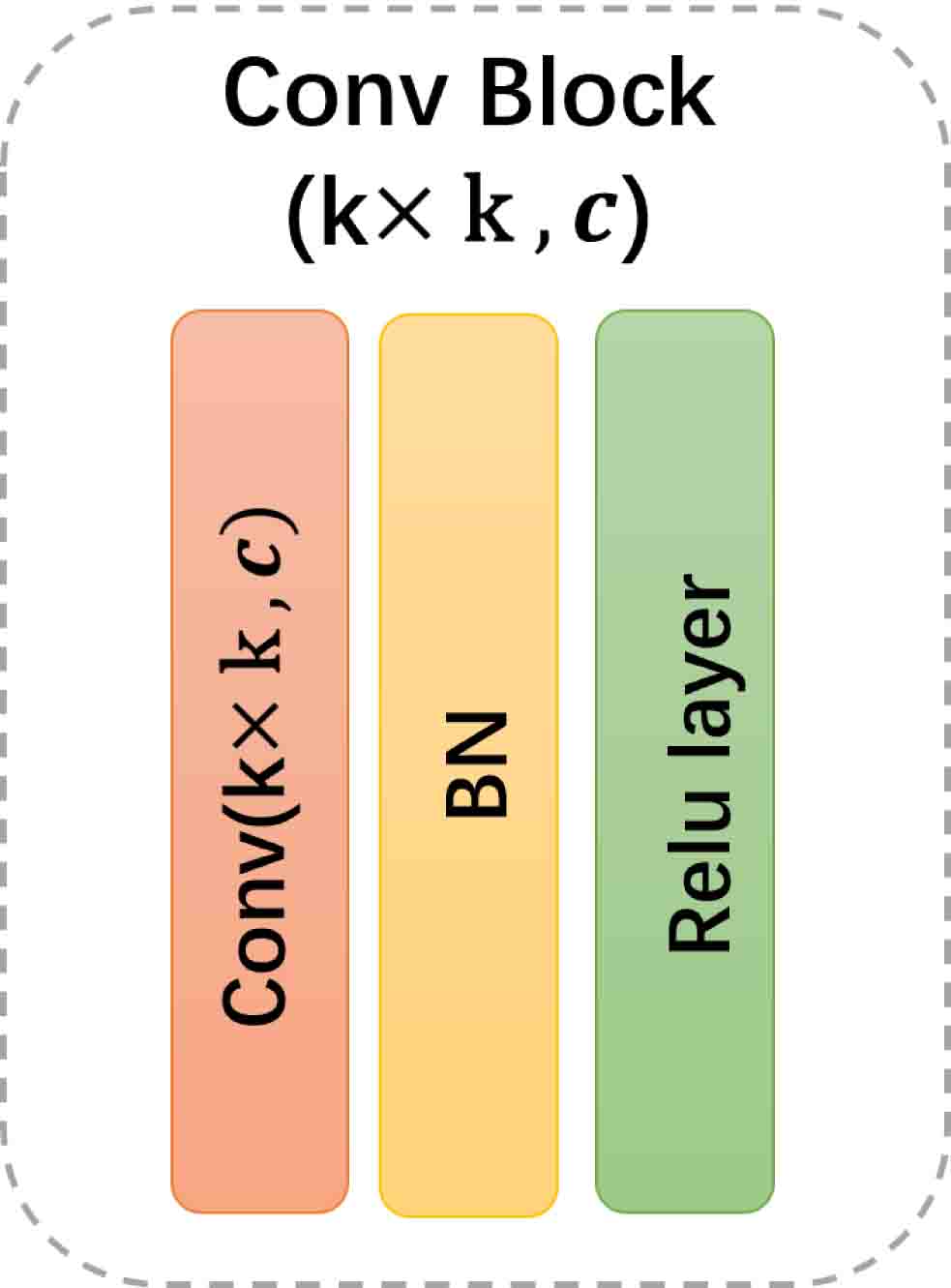}}
	\quad
	\subcaptionbox{Total architecture.\label{sec4_fig:CNN_Coef}}
	{\includegraphics[scale=0.14]{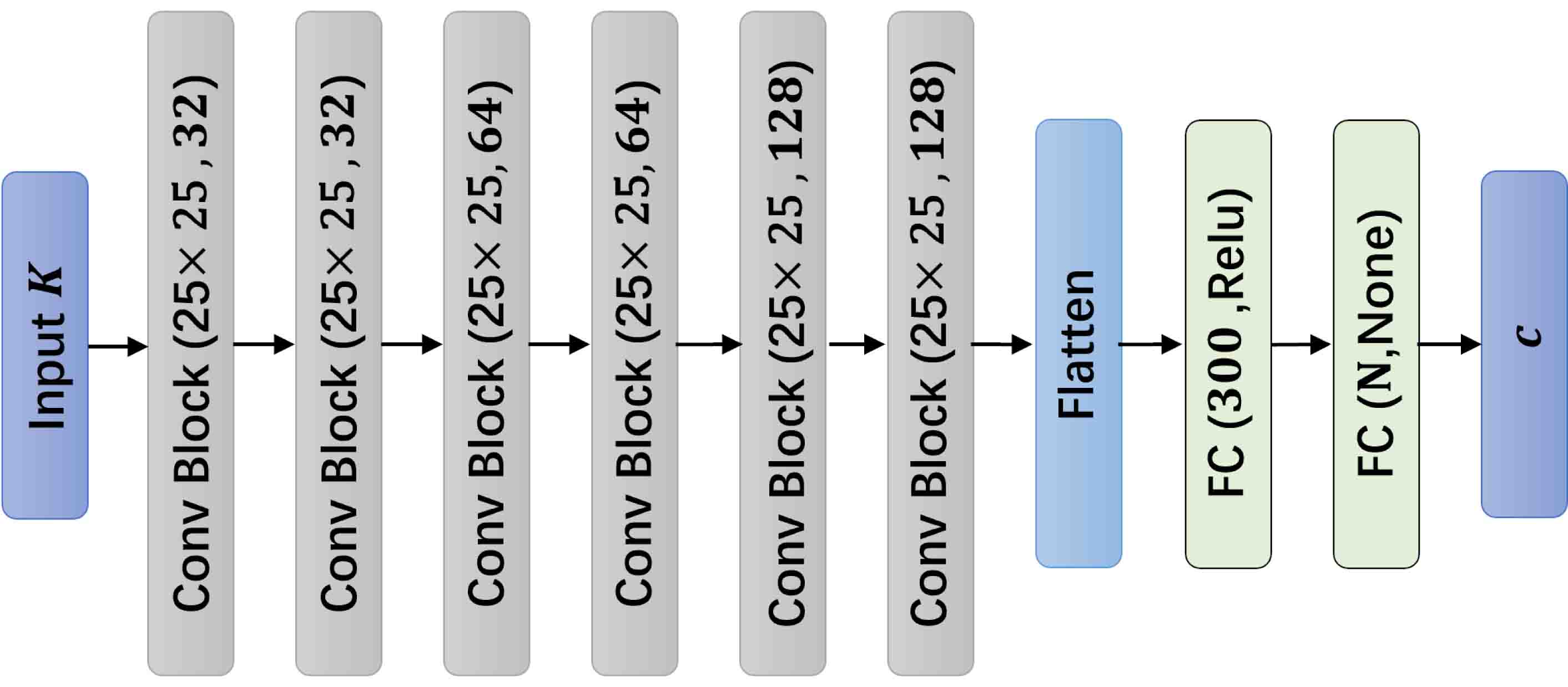}}
	\caption{The architecture of Coef CNNs.}
\end{figure}

The architecture of Coef CNNs is similar to encoders and bottlenecks of Convolutional Autoencoder\cite{sec4:CNN AE}. Convolutional blocks (see Figure \ref{sec4_fig:conv_block}) are first connected to capture local features from input $K$. Fully connected layers (FC) at last few layers help to compress the extracted feature into a smaller vector representation. (see Figure \ref{sec4_fig:CNN_Coef}).

Finally, We use following loss function to train the coefficient CNNs,
\begin{equation}
	\begin{aligned}
		Loss_{Coef}(\phi) = \frac{1}{M}\sum_{i=1}^M\Vert u_h(K_i)-\tilde{u}_h(K_i) \Vert_{L_2}^2.
	\end{aligned}
	\nonumber
\end{equation}

\subsection{Connections with multiscale finite element methods}
\label{ssec4:connections with MsFEM}
\begin{figure}[H]
	\centering
	\includegraphics[scale=0.1]{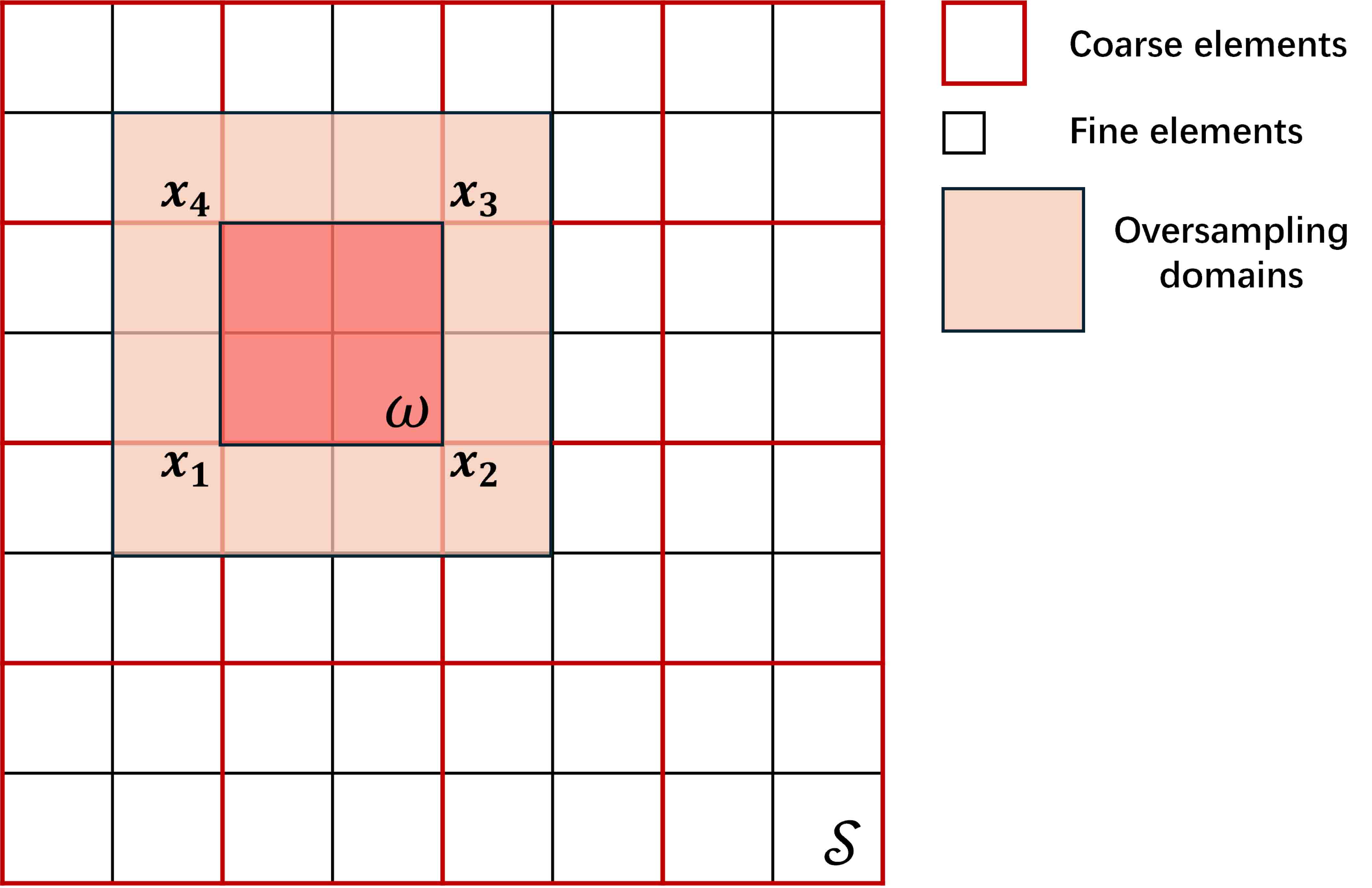}
	\caption{The computational domain of MsFEM.}
	\label{sec4_fig:The computational domain of MsFEM}
\end{figure}

 CNN-based ROM tries to obtain the reduced basis functions by extracting multiscale features from data. Therefore, the Basis CNNs need to be retrained for different source terms. Rather than directly solving above issues, we attempt to find connections between the proposed methods and another class of multiscale model reduction methods, called multiscale finite element methods (MsFEM)\cite{sec1:MsFEM1}. For convenience, we use the diffusion problems as an example (\ref{sec2_eq:diffusion}).

\vspace{0.5em}
The idea that learning input-specific basis functions $P$ by Basis CNNs comes from MsFEM. MsFEM is also a model reduction method, which is realized by constructing multiscale finite element basis functions that are adaptive to the local property of differential operators. Therefore, the basis functions of MsFEM is $K$-specific for the given differential operators. Unlike traditional FEM, which need fine meshes $\mathcal{M}_h$ to capture small-scale features accurately, MsFEM enriches the basis functions by solving local problems on a coarse grid $\mathcal{M}_H$. For each element $\omega \subset \mathcal{M}_H$, four nodal basis functions $\{\Phi_{ms}^i, i=1,2,3,4\}$ satisfy
\begin{equation}
	-\nabla \cdot\Big( \kappa(x,\xi)\nabla \Phi_{ms}^i\Big) =0.
	\label{sec4_eq:cell_problems}
\end{equation}
Let $x_j\in \partial{\omega}$ be the four nodal points of $\omega$. We pose the constraints $\Phi_{ms}^i(x_j)=\delta_{ij}$ and assume that the basis functions vary linearly along the boundary $\partial{S}$. Here, $\delta_{ij}$ is Kronecker Delta fuction.

\vspace{0.5em}
Using CNN-based ROM to learn multiscale finite element basis functions in the oversampling domains. Similar to other numerical upscaling methods, MsFEM also suffers "resonance" effects. Oversampling is a common strategy to overcome this difficulty. In practical, to avoid repeated computations, the spatial space $\mathcal{S}$ is usually decomposed into a number of large sampling regions. Each of these sampling regions contains many coarse elements. The choice of sampling regions size is a balance between accuracy and efficiency. Large size usually leads to faster convergence rate, but it is not practical on the online stage. Here, we use CNN-based ROM to learn the basis functions in a relative large Oversampling domains, which can even be the same as computational domain $\mathcal{S}$. Such application can not only help to reduce online time of MsFEM with oversampling techniques but also breaks some limitations of the proposed method. The implementation details and numerical results can be found in Section \ref{ssec5:Darcy flow in a channelized aquifer}.
\section{Numerical results}
\label{sec5:numerical results}
In this section, the numerical results for various multiscale problems with random inputs are presented to show the accuracy and efficiency of CNN-based ROM method.

\begin{figure}[ht]
	\centering
	\includegraphics[scale=0.12]{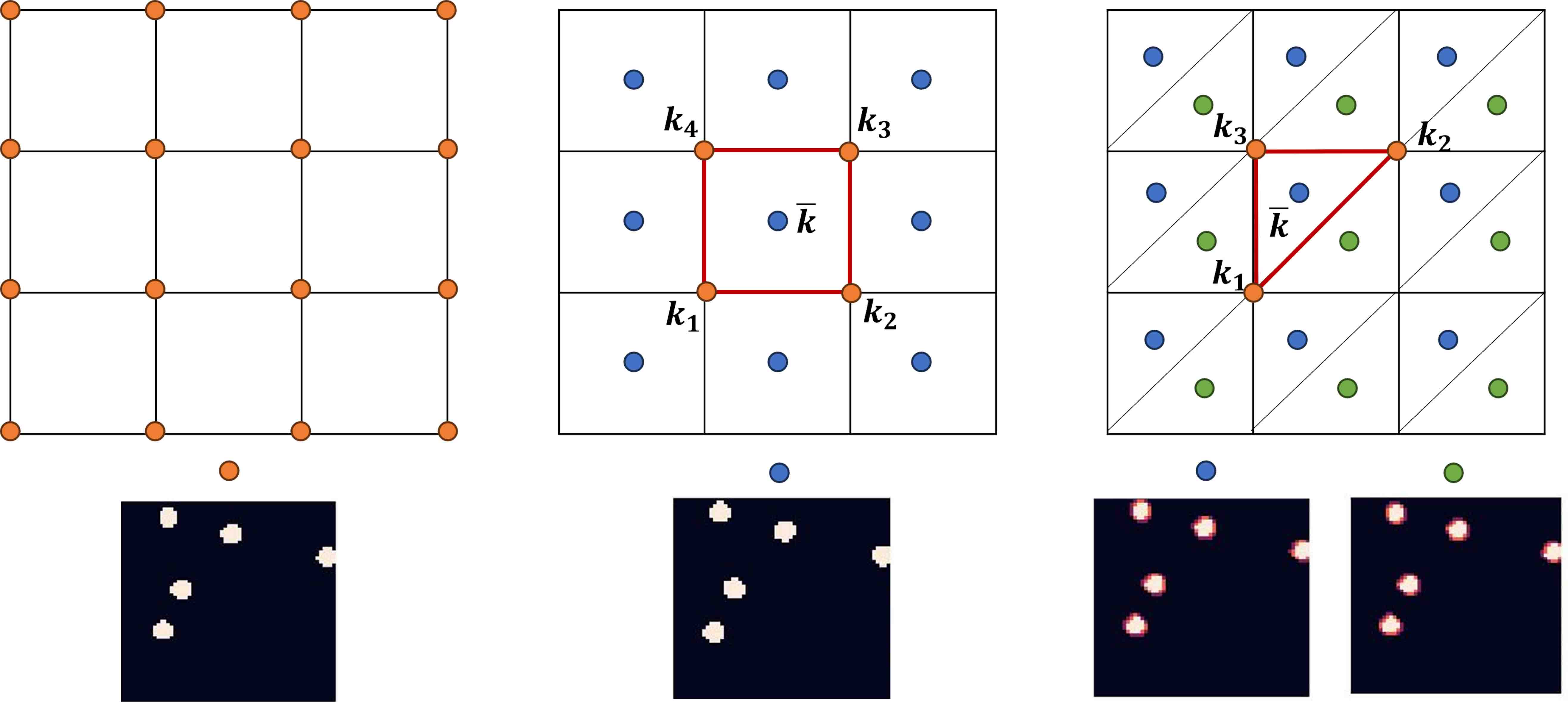}
	\caption{The values of inputs $K$. The points in the figure represent spatial sampling points of $K$. Left: training CNNs. Middle: FEM with rectangular elements. Right: FEM with triangular elements.}
	\label{sec5_fig:grid_nn}
\end{figure}

To illustrate numerical results quantitatively, we define the relative test mean error $\epsilon_{test}$ as follows,
\begin{equation}
	\epsilon_{test} = \frac{1}{M}\sum_{i=1}^M\frac{\Vert u_h(K_i)-u_{CNN}(K_i) \Vert_2^2}{\Vert u_h(K_i) \Vert_2^2},
\end{equation}
where $u_{CNN}$ have different meanings for different cases, specifically, $u_{CNN}=\hat{u}_h$ for Basis CNNs and $u_{CNN}=\tilde{u}_h$ for Coef CNNs. The reference solutions $u_h$ are obtained by FEM methods on mesh $\mathcal{M}$ with either square or triangle elements. The inputs $K_i$ of both neural networks are the samples of $K$ defined in Section \ref{sec2:problem setup}. It should be noted that FEM methods require the value of $\kappa$ on the integral points while we now only have the values on the nodes. To deal with this contradiction, we use element means for the integral points (see Figure \ref{sec5_fig:grid_nn}).

\subsection{Applications to Darcy flow}
\label{ssec5:Application to Darcy flow}
The Darcy flow, which models how flow moves in a porous medium, has been widely studied in civil, geotechnical, and petroleum engineering \cite{sec5:DarcyFlow}. In this subsection, we consider the single-phase Darcy flow for pressure $u(x)$, i.e.,
\begin{equation}
	\left\{
		\begin{aligned}
			&-\nabla \cdot\Big( \kappa(x,\xi)u(x)\Big) = 1, x \in \mathcal{S}:=[0,1]\times[0,1],\\
			&u(x)=0, x \in \partial{\mathcal{S}}.
		\end{aligned}
		\right.
	\label{sec5_eq:exper1_equation}
\end{equation}
The weak form of above equation is the following:
\begin{equation}
	\text{Find} \ u, u\in V, \text{such that},
	\int_{\mathcal{S}}\kappa \nabla u \cdot \nabla v dx = \int_{\mathcal{S}} v dx, \forall v\in V.
	\nonumber
\end{equation}
In this section, we use finite-dimensional space $V_h$ with bilinear basis functions $\varphi_1,\ldots,\varphi_{N_h^2}$ defined on rectangular elements to obtain the high-fidelity approximation of $u$, then the stiffness matrix $A_h(K)$ belongs to $\mathbb{R}^{N_h^2\times N_h^2}$ and has entries
\begin{equation}
	a_h(i,j)=\int_{\mathcal{S}}K \nabla \varphi_i \cdot \nabla \varphi_j dx,
	\label{sec5_eq:exper1_aij}
\end{equation}
and the load vector $F_h(K)$ belongs to $\mathbb{R}^{N_h^2\times 1}$ and has entries
\begin{equation}
	f_h(i)=\int_{\mathcal{S}} \varphi_i dx.
	\label{sec5_eq:exper1_fi}
\end{equation}
The approximated solution can thus be obtained by
\begin{equation}
	u_h(K) = A_h(K)^{-1}F_h(K).
	\label{sec5_eq:exper1_uh}
\end{equation}

Next, we will show that the proposed methods works for different types of random inputs $\kappa$. In addition, since Darcy flow (\ref{sec5_eq:exper1_equation}) is a linear equation, we only use Basis CNNs to learn the mapping from $K$ to $uh$ for convenience. The efficacy of Coef CNNs will be demonstrated in Section \ref{ssec5:Application to nonlinear flows}.

\subsubsection{Darcy flow with binomial point process}
\label{ssec5:Darcy flow with binomial point process}
\begin{figure}[H]
	\centering
	\includegraphics[scale=0.23]{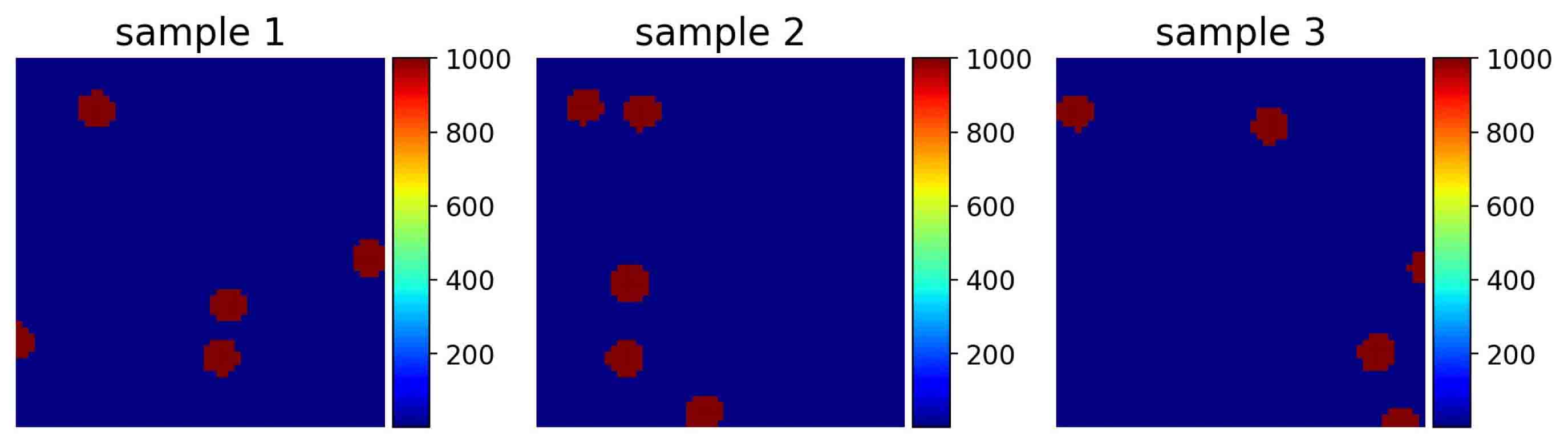}
	\caption{Three samples of binomial point process defined in Section \ref{ssec5:Darcy flow with binomial point process}.}
	\label{sec5_fig:exper1_K}
\end{figure}

In the fist test case, we consider the binomial point process $\kappa(x,\xi)$ defined on $\mathcal{P}=(\Omega,\mathcal{F},\mathbb{P})$ for any given $x\in \mathcal{S}$. Let any subset of $\mathcal{S}$ be $\mathcal{B}$ and $\{X_i\}_{i=1}^n$ be n $i.i.d$ points with the distribution $\mathbb{P}(\mathcal{S})$ and the density $p(x)$. Then the number of points in $\mathcal{\mathcal{B}}$ can be defined as the following random variable
\begin{equation}
	\xi(\mathcal{B}) = \sum_{i=1}^n \delta(X_i\in \mathcal{B}),
\nonumber
\end{equation}
where $\delta(X_i\in \mathcal{B})$ is the indicator function and takes value as
\begin{equation}
	\delta(X_i\in \mathcal{B})=
	\left\{
		\begin{aligned}
			&1\ \ \ \mathrm{if}\ X_i\in \mathcal{B},\\
            &0\ \ \ \mathrm{otherwise}.
		\end{aligned}
		\right.
	\nonumber
\end{equation}
The probability of each random variables $X_i$ in $\mathcal{B}$ is the same and can be defined as follows,
\begin{equation}
	p = \mathbb{P}(X_i\in \mathcal{B})=\frac{\int_{\mathcal{B}}p(x)dx}{\int_{\mathcal{S}}p(x)dx}.
	\nonumber
\end{equation}
Then $\xi(\mathcal{B})$ is distributed to a binomial distribution with $n$ and $p$. We introduce the set
\begin{equation}
	F(\xi) = \cup_{i=1}^n \{x\in\mathbb{R}^2| \Vert x- x_i\Vert_2\le r\},
	\nonumber
\end{equation}
where $x_i$ is the sample of $X_i\sim p(\mathcal{S})$ and $r$ is the radius of points. Then we can define the two-scale random permeability $\kappa(x,\xi)$, where points of permeability $\kappa^{(1)}$ are randomly distributed in the background media of conductivity $\kappa^{(0)}$. Specifically,
\begin{equation}
	\kappa(x,\xi)=
	\left\{
		\begin{aligned}
			&\kappa^{(1)}\ \ \  \mathrm{if}\ x\in F(\xi),\\
            &\kappa^{(0)}\ \ \  \mathrm{otherwise}.
		\end{aligned}
		\right.
	\nonumber
\end{equation}
Here, we consider the special case that $\mathcal{B}=\mathcal{S}$, $\mathbb{P}$ is the uniform distribution on $\mathcal{S}$ and takes $n=5, r=0.05, \kappa^{(1)}=1000, \kappa^{(0)}=1$. In the following experiments, we discretize $\kappa(x,\xi)$ over the uniform FEM mesh $\mathcal{M}$ with rectangular elements and $61 \times 61$ nodes, denoted by $K$. Several instances of $K$ are given in Figure \ref{sec5_fig:exper1_K}.

In this paper, we only approximate the solutions on the free nodes, i.e., the spatial resolution of inputs is $59\times 59$. The configuration of Basis CNNs are shown in Figure \ref{sec4_fig:CNN_Base} with more details in Table \ref{sec5_table:architecture of Basis CNN}. The 2nd-3th columns of Table 1 show the spatial resolution $H\times W$ of feature maps and the number of parameters of each layer. This configuration will be used in all experiments except that the spatial resolution of feature maps varies with different cases.
\begin{table}[H]
	\centering  
	\setlength{\abovecaptionskip}{0cm}
	\setlength{\belowcaptionskip}{0.2cm}
	\caption{The architecture of Basis CNNs}  
	\label{sec5_table:architecture of Basis CNN}  
	\scalebox{0.8}{
		\begin{tabular}{ccc}
			\toprule  
			  Layers &Resolution H $\times$ W & The number of parameters \\  
			\midrule
			& & \\[-6pt]  
			Input &59$\times$ 59 &-\\
			& & \\[-6pt]  
			Conv$\_$p block (25$\times$ 25,32,1) &59$\times$ 59 &20000 \\
			& & \\[-6pt]  
			Conv$\_$p block (25$\times$ 25,32,1) &59$\times$ 59 &20000 \\
			& & \\[-6pt]  
			Conv$\_$p block (25$\times$ 25,64,1) &59$\times$ 59 &40000\\
			& & \\[-6pt]  
			Conv$\_$p block (25$\times$ 25,64,1) &59$\times$ 59 &40000\\
			& & \\[-6pt]  
			Conv$\_$p block (25$\times$ 25,128,1) &59$\times$ 59 &80000\\
			& & \\[-6pt]  
			Conv$\_$p block (25$\times$ 25,128,1) &59$\times$ 59 &80000\\
			& & \\[-6pt]  
			Conv(25$\times$ 25,1,1) &59$\times$ 59 &625\\
			& & \\[-6pt]  
			Reshape &3481 $\times$ 10 &-\\
			\bottomrule
		\end{tabular}
	}
\end{table}
 The Basis CNNs are trained with Adam optimizer with loss (\ref{sec4_eq:loss function(Base)}). The initial learning rate is 0.0001 and cosine decay, which is a built-in function of TensorFlow, is used. The batch size is set as 32 and the model is trained for 100 epochs. Each sample in the dataset is composed of four variables, they are permeability $K$, high-fidelity solution $u_h(K)$, stiffness matrix $A_h(K)$ and load vector $F_h(K)$, respectively (see equation (\ref{sec5_eq:exper1_aij})-(\ref{sec5_eq:exper1_uh})). We collect the data pairs for 12740 different samples of $K$, 10240 of which are selected as training data and the rest are used to test the generalization ability of CNN-based ROM.

\begin{figure}[ht]
	\centering
	\includegraphics[scale=0.23]{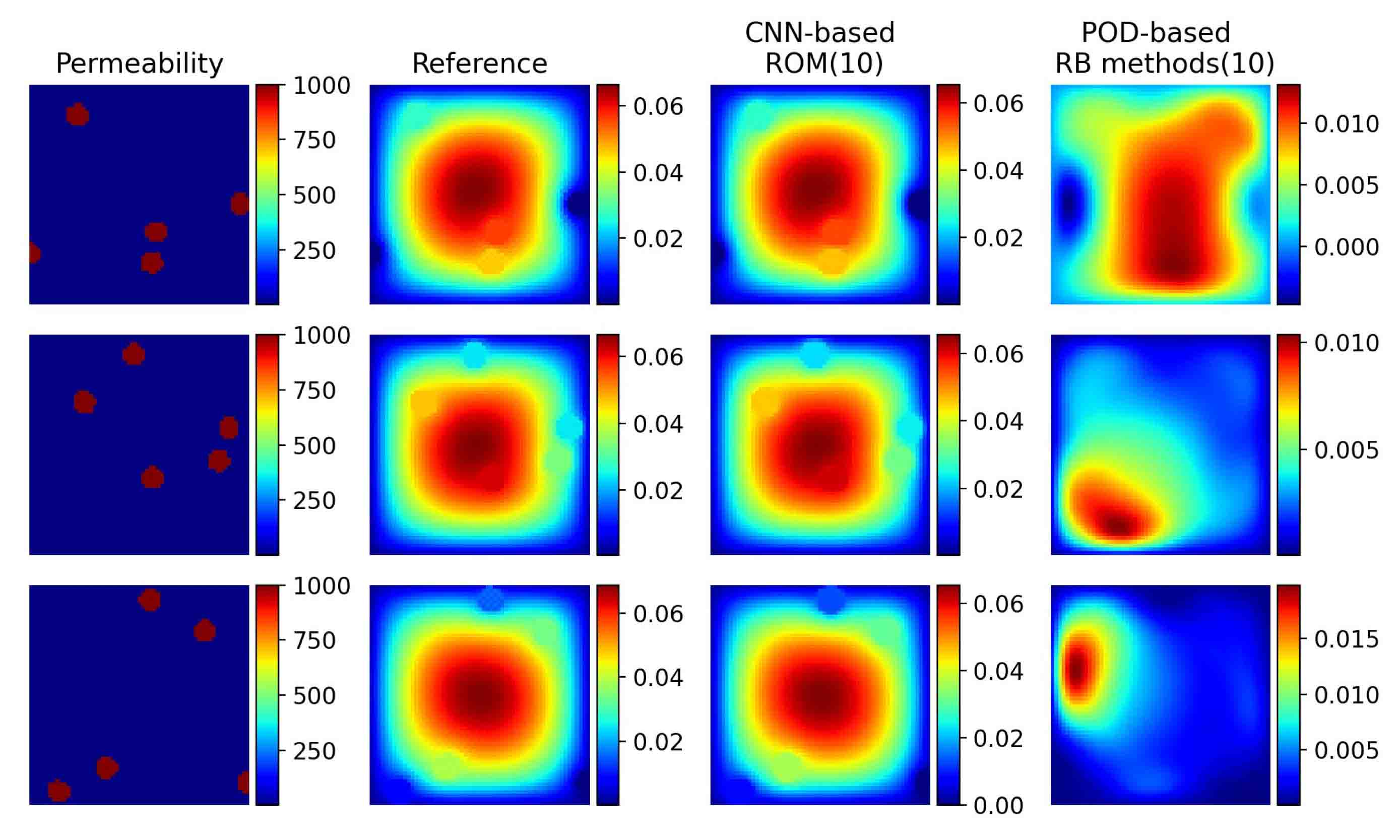}
	\caption{the predictive results of CNN-based ROM and POD-based RB methods with ten basis functions at three test instances of $K$.}
	\label{sec5_fig:exper1_comparasion_withoutnoise}
\end{figure}

\begin{figure}[ht]
	\centering
	\includegraphics[scale=0.19]{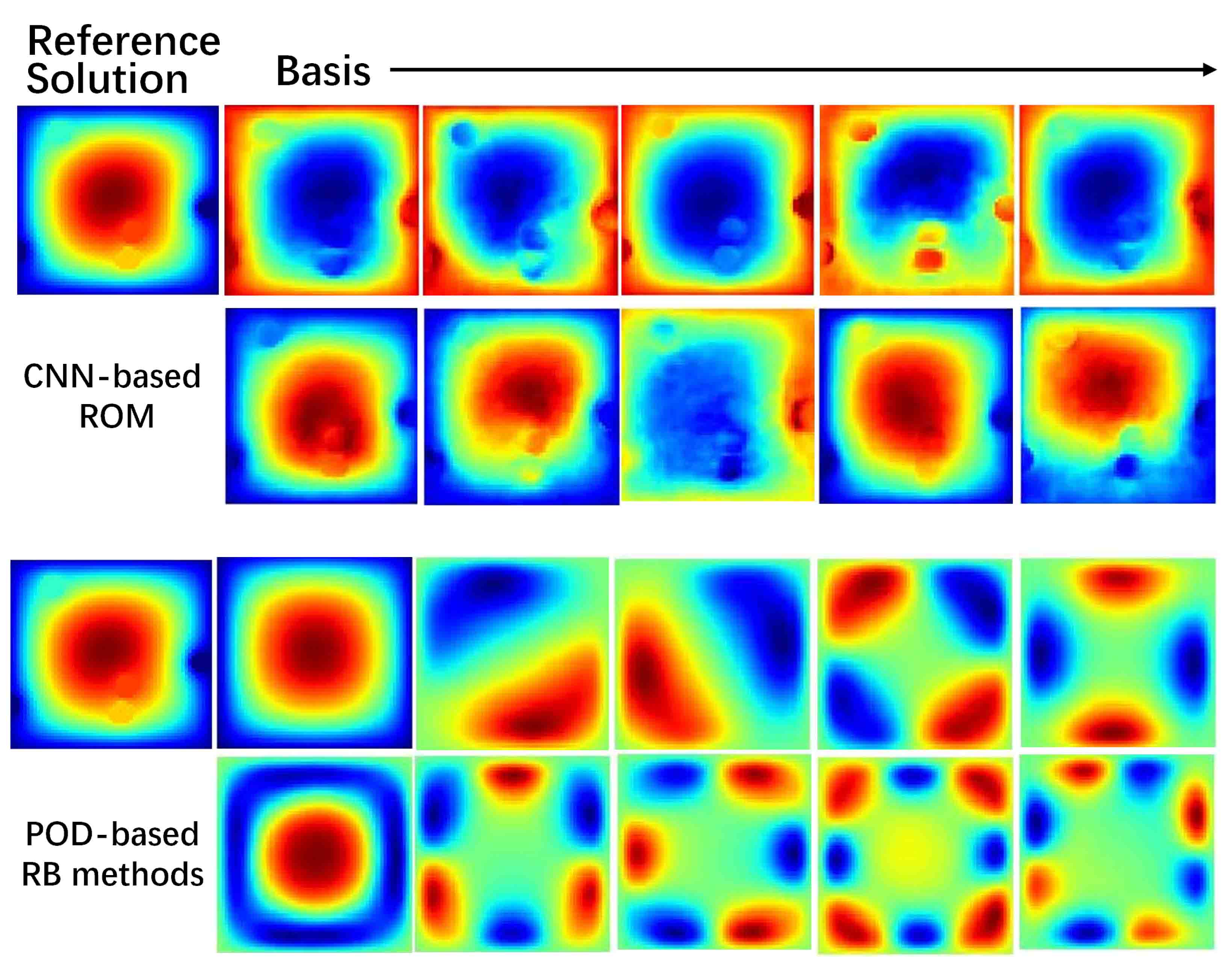}
	\caption{Basis functions for test instance in the first row of Figure \ref{sec5_fig:exper1_comparasion_withoutnoise}. The first two lines show the basis functions of CNN-based ROM with $N=10$. Ten POD basis functions corresponding to the ten largest eigenvalues are listed in the last two rows. }
	\label{sec5_fig:exper1_basis}
\end{figure}

 \vspace{0.5em}
 \noindent\textbf{Compared with POD-based RB methods.} In the following context, we compare the predictive performance of CNN-based ROM with POD-based RB methods introduced in Section \ref{sec3:Preliminary}. Figure \ref{sec5_fig:exper1_comparasion_withoutnoise} illustrates that CNN-based ROM can approximate the high-fidelity solution accurately using few number of RB basis functions ($N=10$), which can not be achieved by POD-based RB methods. Since the accuracy of POD-based RB methods is highly dependent on the number of RB basis functions $N$, we let $N$ equal 10, 20, 30, 40, 50, 800, 900, 1000, respectively and record the relative test mean errors in Table \ref{sec5_exper1:the affect of the number of basis}. As shown in table, POD-based RB methods requires 900-1000 RB basis functions to achieve the same accuracy as the proposed method, which directly affects the efficiency of the online computation.
\begin{table}[H]
\centering  
\setlength{\abovecaptionskip}{0cm}
\setlength{\belowcaptionskip}{0.2cm}
\caption{Darcy flow with binomial point process: comparison of relative test mean errors for different basis numbers between CNN-based ROM and POD-based RB method. }  
\label{sec5_exper1:the affect of the number of basis}  
\scalebox{0.8}{
	\begin{tabular}{ccccccccc}
		\toprule  
		The number of basis	& 10 & 20 &30 &40 &50 &800 &900 &1000 \\  
		\midrule
		& & & & & & & & \\[-6pt]  
		POD-based RB methods &81.42\% &65.56\% &54.50\% &47.96\% &43.47\% &4.46\% &3.68\% &3.08\% \\
		& & & & & & & & \\[-6pt]  
		CNN-based ROM &3.52\% & 3.39\% &3.29\% & 3.23\% &3.43\% &- &- &- \\
		\bottomrule
	\end{tabular}
}
\end{table}
The RB methods exploit low-rank features of the data, which is usually called modes, that characterize physical processes. Therefore, we show the RB basis functions learned by CNN-based ROM and POD-based RB methods in Figure \ref{sec5_fig:exper1_basis}. Note that for POD-based RB basis functions, the modes are distributed around the center and dominant modes alternate arrangement. In contrast, the shape of the CNN-based RB basis functions are very much like the data. Such input-different basis functions can help reduce the number of basis functions needed for approximation. However, they lack interpretability, which is the common problem of deep learning techniques.

\begin{figure}[ht]
	\centering
	\includegraphics[scale=0.23]{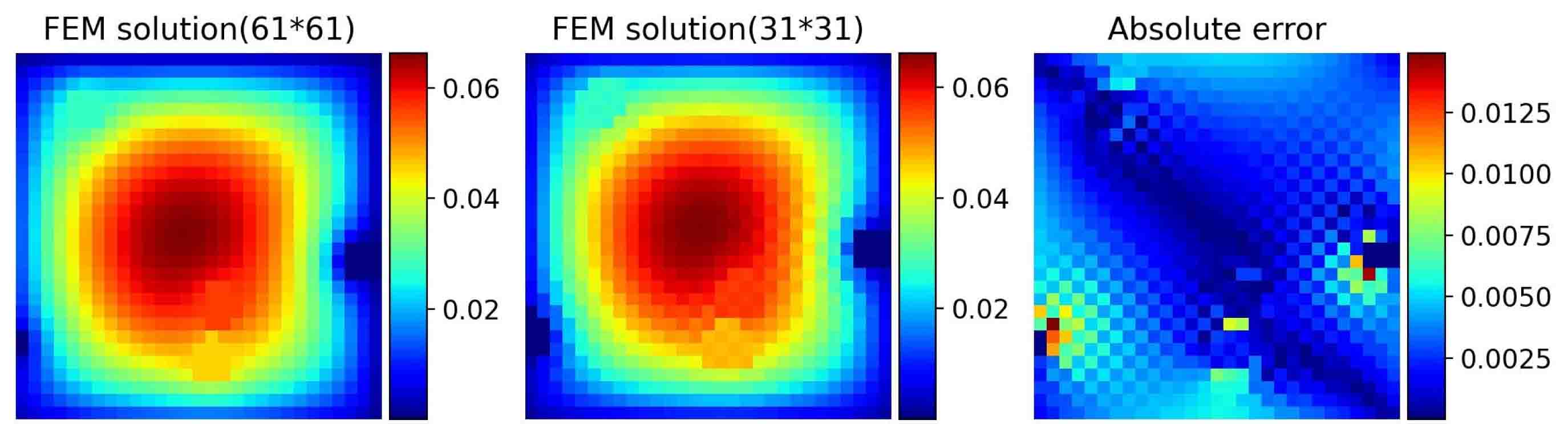}
	\caption{From left to right are FEM solutions on mesh with $61\times 61$ nodes, FEM solutions on mesh with $31\times 31$ nodes and their absolute error, respectively.}
	\label{sec5_fig:exper1_coarse_solution}
\end{figure}

\begin{figure}[ht]
	\centering
	\includegraphics[scale=0.23]{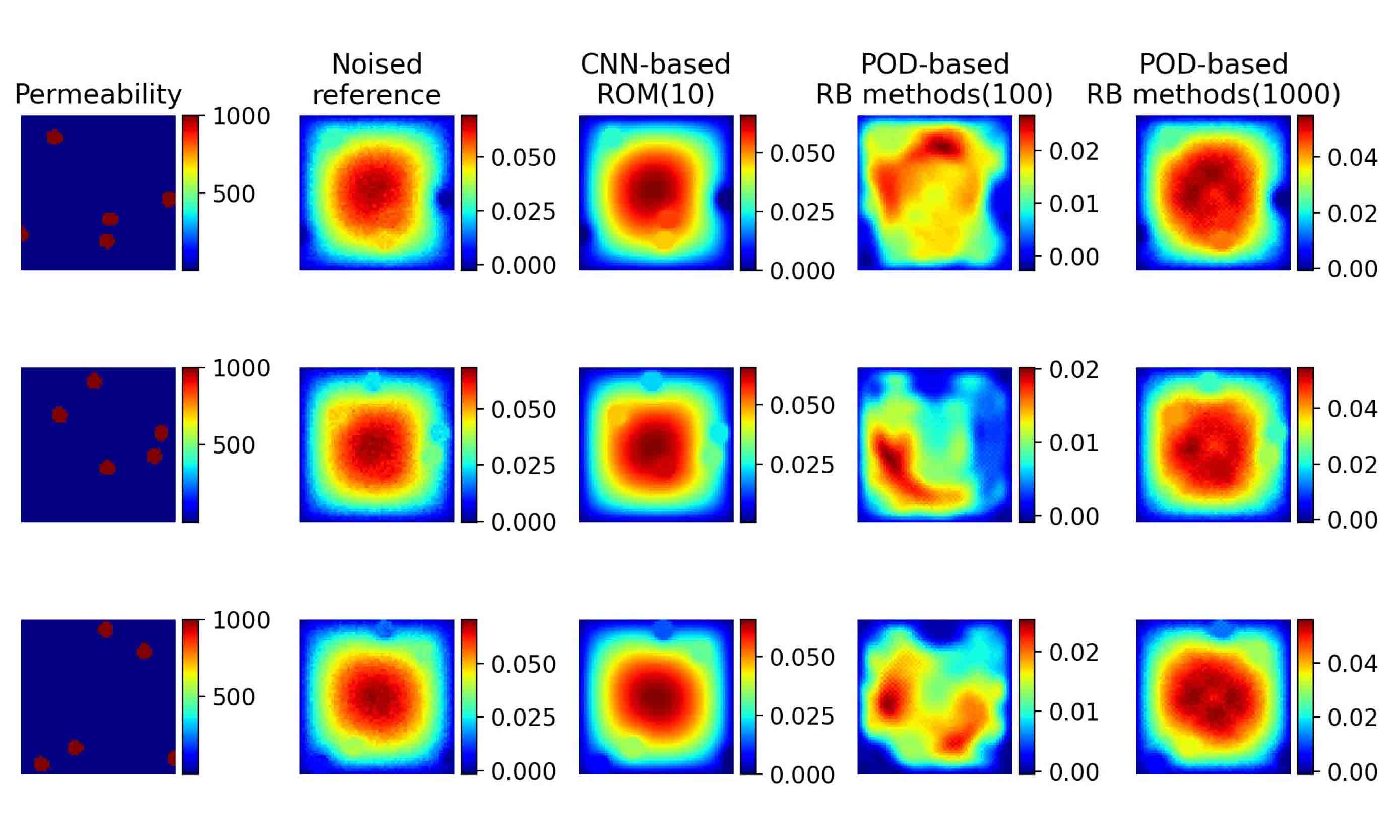}
	\caption{The predictive results for data with Gaussian noise at a level of 0.001.}
	\label{sec5_fig:exper1_comparasion_withnoise}
\end{figure}

Our training data are generated by FEM, which has approximation error. For instance, FEM solutions on a mesh with $31\times 31$ nodes are less accurate than those on a mesh with $61\times 61$ (see Figure \ref{sec5_fig:exper1_coarse_solution}). We use a Gaussian distribution with means of zero and standard deviations of 0.001 to simulate the errors caused by numerical methods, which is shown in the second column of Figure \ref{sec5_fig:exper1_comparasion_withnoise}.
Using the noisy data, we train the Basis CNNs with $N=10$ and implement POD-based RB methods with $N=100,1000$ again. The numerical results are presented in Figure \ref{sec5_fig:exper1_comparasion_withnoise}. CNN-based ROM shows a clear advantage and a de-noising effect.

\subsubsection{Darcy flow in a channelized aquifer}
\label{ssec5:Darcy flow in a channelized aquifer}
\begin{figure}[ht]
	\centering
	\includegraphics[scale=0.13]{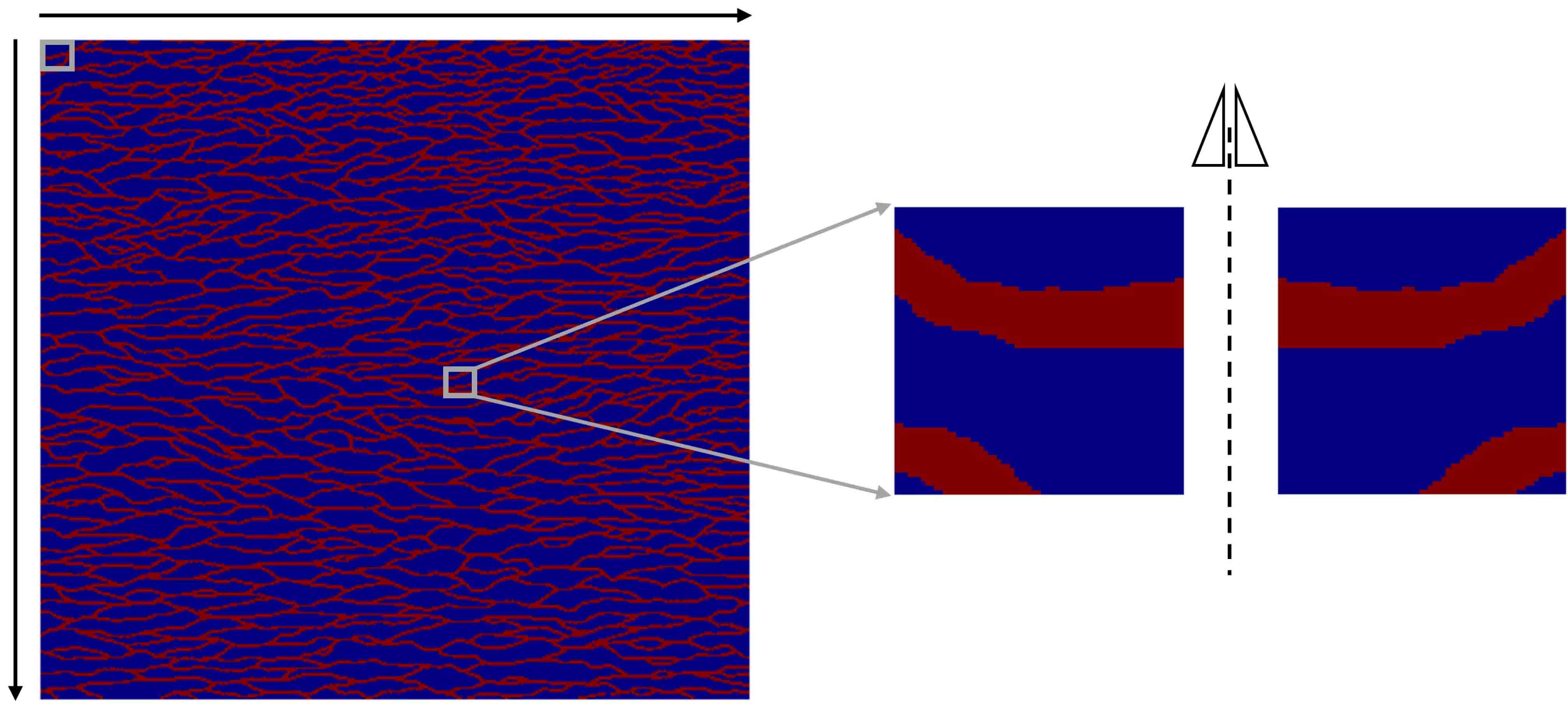}
	\caption{The hydraulic conductivity field of a channelized aquifer.}
	\label{sec5_fig:exper2_background}
\end{figure}

\begin{figure}[ht]
	\centering
	\includegraphics[scale=0.23]{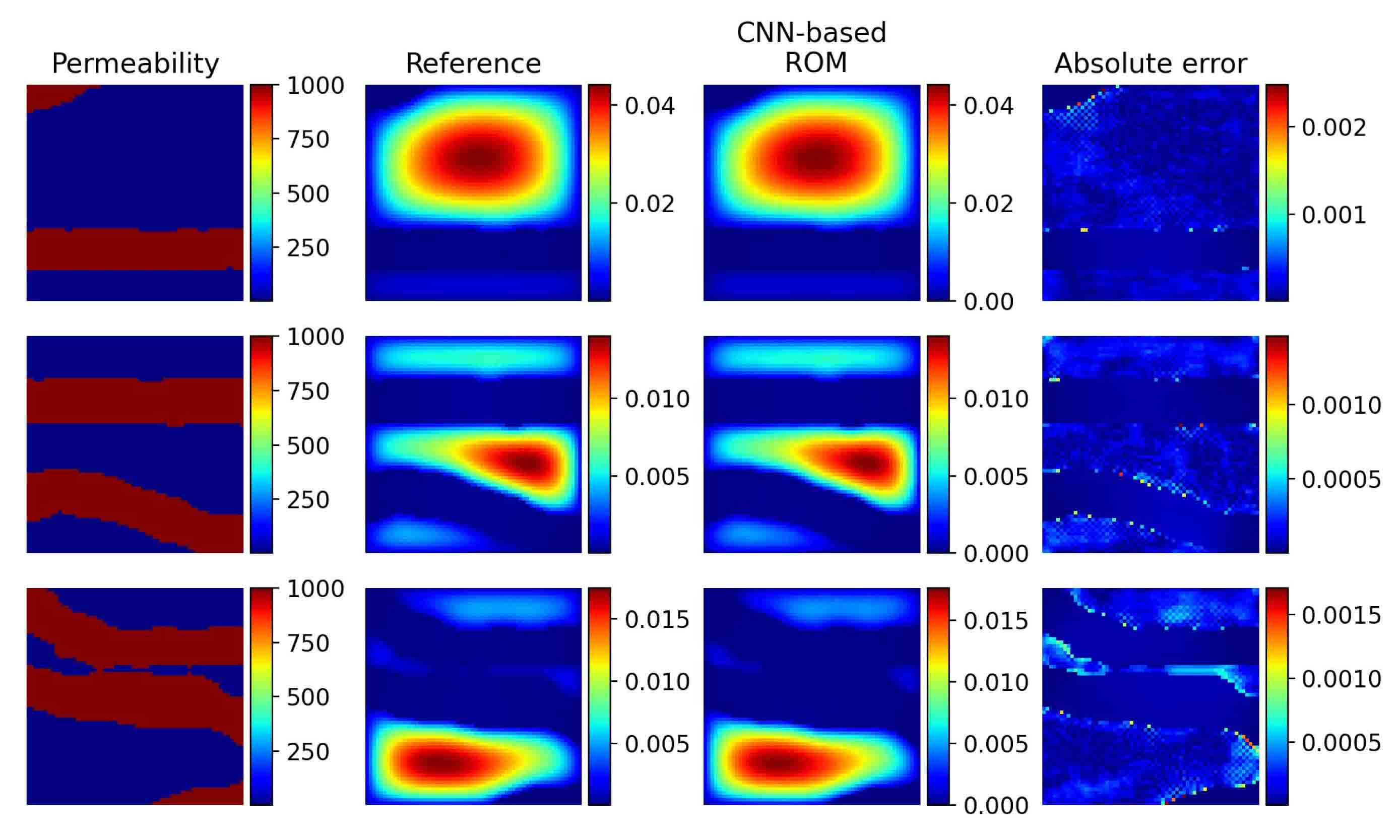}
	\caption{The predictive results of CNN-based ROM with ten basis functions at three test instances of $K$.}
	\label{sec5_fig:exper2_solution}
\end{figure}

In this subsection, we consider the random input in Figure \ref{sec5_fig:exper2_background}, which simulates the hydraulic conductivity field of a channelized aquifer \cite{sec5:groundwater}. The total number of pixels is $2500\times 2500$. Channel and matrix materials are assigned hydraulic conductivity values of 1000 and 1, respectively. To introduce randomness, we extract 23104 small images of resolution $64\times 64$ from the large image by moving along both the horizontal and vertical directions with an interval of 16 pixels for each movement. To obtain enough training data, we flip the generated images horizontally (see Figure \ref{sec5_fig:exper2_background}). We finally get 46208 images, among which 40960 images are collected as training dataset and the rest are used for test. Then Basis CNNs with architecture (Table \ref{sec5_table:architecture of Basis CNN}) are trained using Adam optimizer for 50 epochs. The relative tests mean error is $2.16\%$ and the generalization ability of the proposed method is intuitively presented in Figure \ref{sec5_fig:exper2_solution}.

Next, we use the upper-left quarter of this background field to validate the ideas in Section \ref{ssec4:connections with MsFEM}, i.e., the total number of pixels is $1250 \times 1250$. Here, we consider an extreme case that the oversampling domain covers the whole computational region. Linear boundary conditions defined in equation (\ref{sec4_eq:cell_problems}) are used for the oversampling domain. Four nodal basis functions are obtained with different boundary conditions. Rather than training four different CNNs, we fully utilize the rotational symmetry of the boundary conditions, i.e., each boundary condition can be rotated clockwise by 90 degrees to overlap with another boundary condition. This can be accomplished by rotating background field. More specifically, we use Basis CNNs to learn the following equation,
\begin{equation}
	\left\{
	\begin{aligned}
	&-\nabla \cdot\Big( \kappa(x,\xi)\nabla \tilde{\Phi}_{ms}(x)\Big) = 0,\ x\in \mathcal{S}:=[0,1]\times[0,1],\\
	&\tilde{\Phi}_{ms}(0,1) = 1,  \tilde{\Phi}(0,0)=0, \tilde{\Phi}_{ms}(1,0)=0, \tilde{\Phi}_{ms}(1,1)=0.
	\end{aligned}
	\right.
	\nonumber
\end{equation}

\begin{figure}[ht]
	\centering
	\includegraphics[scale=0.13]{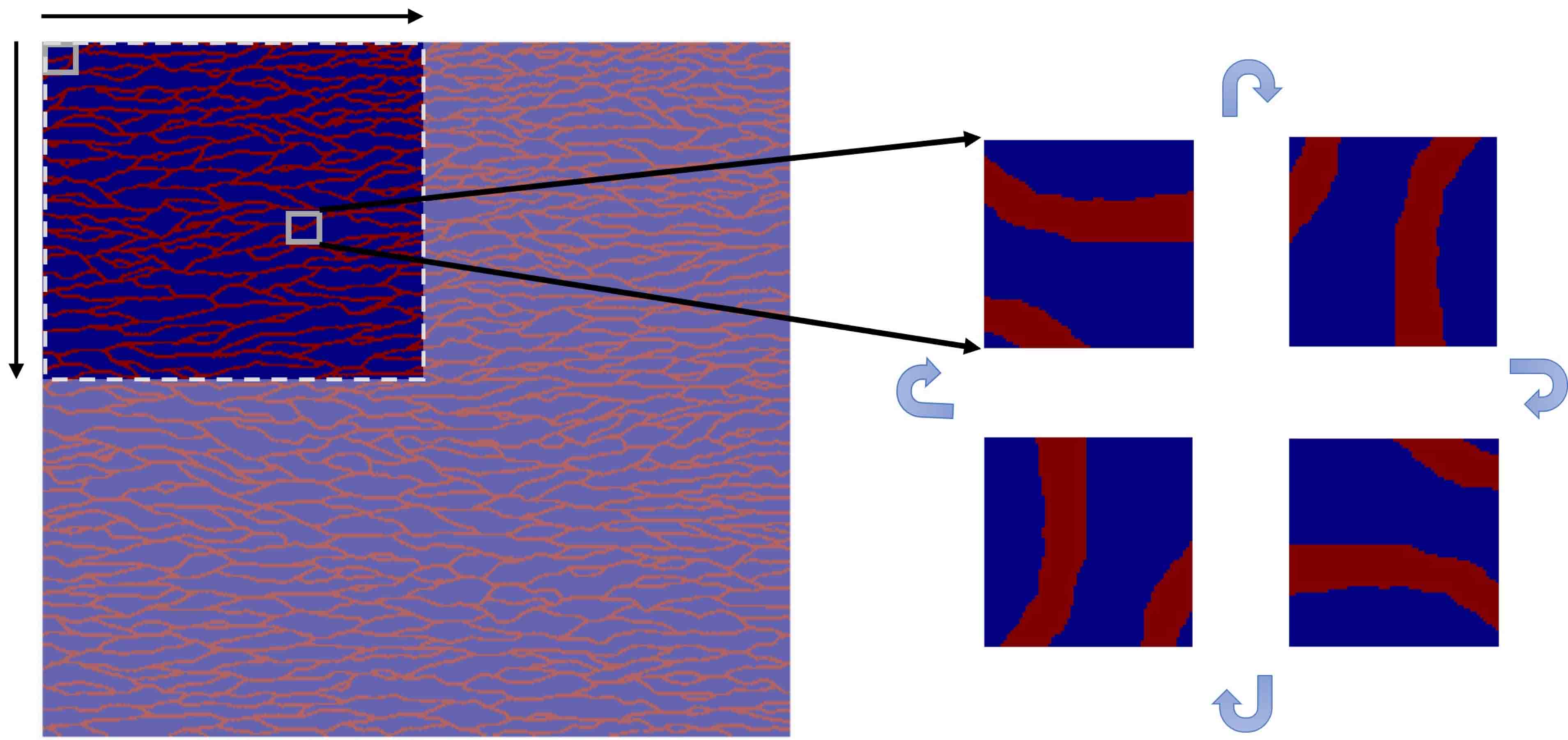}
	\caption{The hydraulic conductivity field of a channelized aquifer for learning multiscale finite element basis functions.}
	\label{sec5_fig:exper2_background2}
\end{figure}

To generate training and test dataset, we first obtain 5625 sub-image of resolution $65\times 65$ by moving along both the horizontal and vertical directions with an interval of 16 pixels for each movement. Then each sample are rotated clockwise by 90 degrees three times to acquire three additional samples (see Figure \ref{sec5_fig:exper2_background2}).

\begin{figure}[ht]
	\centering
	\includegraphics[scale=0.27]{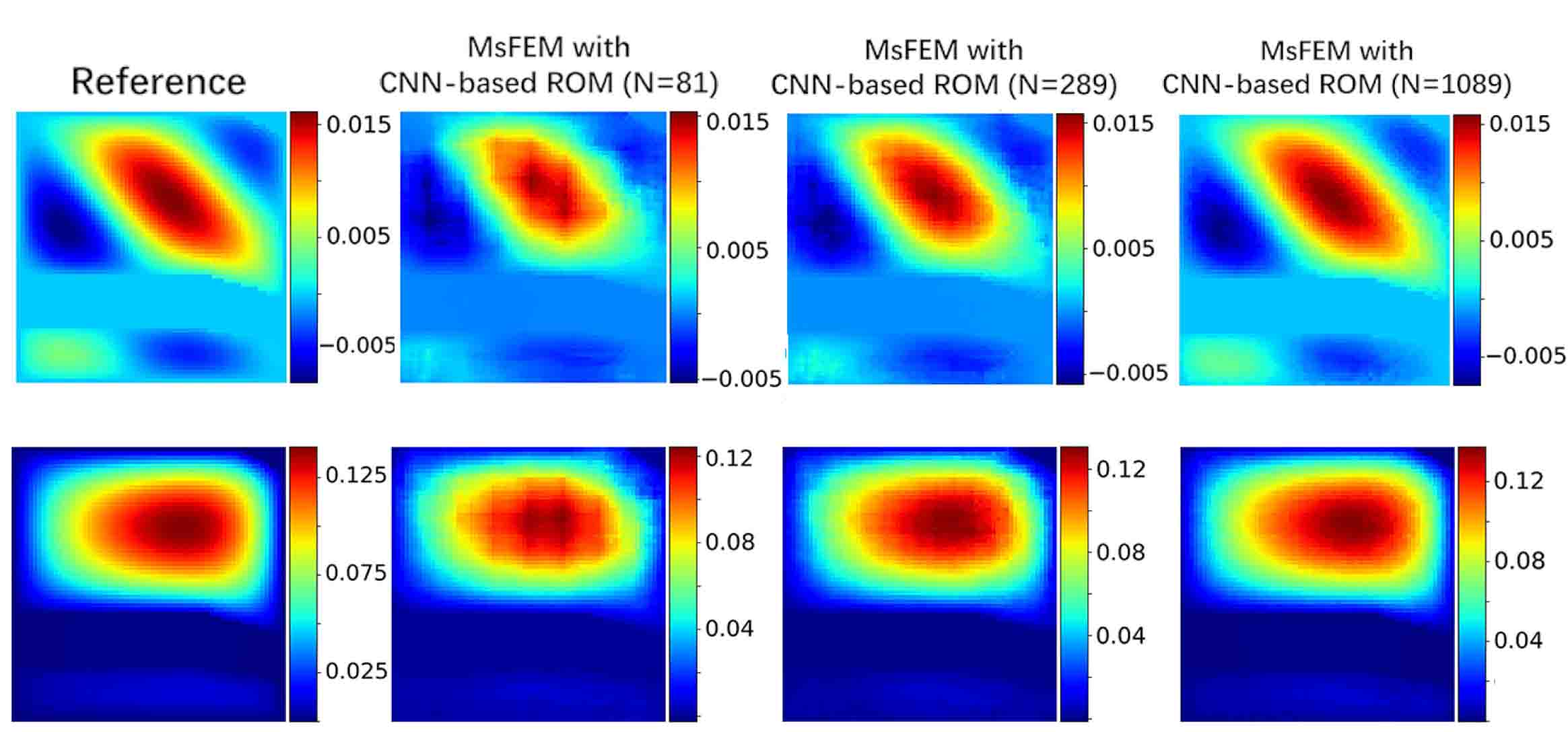}
	\caption{MsFEM with multiscale basis functions learned by CNN-based ROM for different number of basis functions $N=81, 289, 1089$. The source terms for each rows in the figure are $f(x)= \exp(x_1+x_2)$ and $f(x)= \sin(2\pi x_1+2\pi x_2)$, respectively.}
	\label{sec5_fig:exper2_msfem_righthand}
\end{figure}

The relative test mean error is $1.38\%$ and several test instances are shown in Figure \ref{sec5_fig:exper2_msfem_righthand}. Note that the basis functions learned by Basis CNNs are temporary, we construct the true multiscale finite element basis functions $\Phi_{ms}^i, i=1,2,3,4$ in each coarse elements $\omega$ from the linear combination of $\tilde{\Phi}_{ms}^i, i=1,2,3,4$, i.e.,
\begin{equation}
	\Phi_{ms}^i = \sum_{k=1}^4 c_{ik}\tilde{\Phi}_{ms}^k,
	\nonumber
\end{equation}
where the coefficients $c_{ik}$ is determined by conditions $\Phi_{ms}^i(x_j)=\delta_{ij}$. This leads to $N$ non-conforming basis functions which is dealt with by simply taking averages on the boundary because there only exists an $O(1)$ jump across $\partial{\omega}$ \cite{sec1:MsFEM1}. The numerical results depicted in Figure \ref{sec5_fig:exper2_msfem_righthand} demonstrate that the same basis functions can be used for different types of source terms and the numerical solution is convergent to the reference as the number of basis functions is large enough.

\subsection{Applications to nonlinear flows}
\label{ssec5:Application to nonlinear flows}
\begin{figure}[ht]
	\centering
	\includegraphics[scale=0.4]{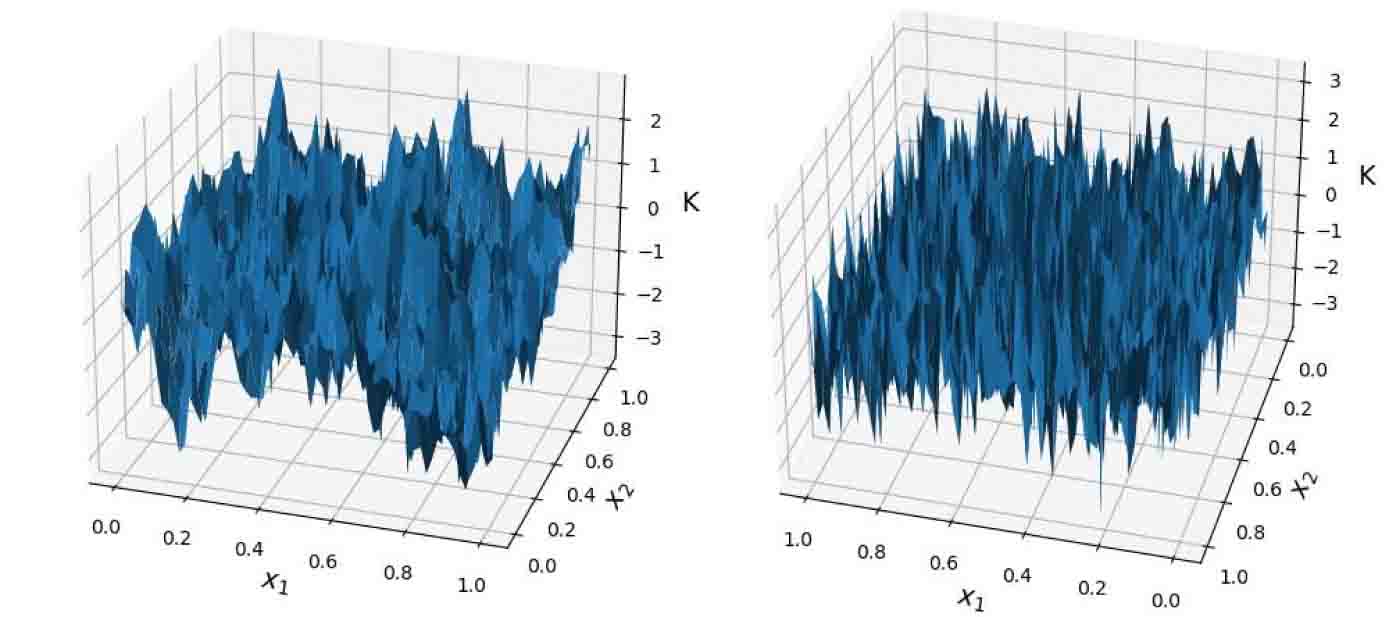}
	\caption{Gaussian random field with bandwidth 0.1 (left) and 0.01 (right).}
	\label{sec5_fig:exper3_K}
\end{figure}

In this subsection, we consider more complex cases, namely nonlinear flows. The logarithm of the random input $\kappa(x,\xi)$ is restricted to be a Gaussian random field, i.e.,
\begin{equation}
	\kappa(x,\xi) = \exp \Big(\xi(x)\Big), \xi(\cdot)\sim \mathcal{N}\Big(m(\cdot),k(\cdot,\cdot)\Big),
\end{equation}
where $m:\mathcal{S}\rightarrow \mathbb{R}$ is the constant mean function with value $m$. The covariance function $k(\cdot,\cdot)$ is a finite positive semi-definite function, which is also call reproducing kernel on $\mathcal{S}$ \cite{sec1:Gaussian Field}. In this test case, we take $m=0$ and the covariance function $k$ is specified in the following form using vector 2-norm,
\begin{equation}
	k(x_1,x_2)=\exp \Big(-\frac{\Vert x_1-x_2\Vert_2}{l}\Big),
\end{equation}
where $l$ is a hyperparameter called bandwidth. Decreasing the bandwidth leads to the covariance between points $x_1$ and $x_2$ decreasing at a faster rate with respect to Euclidean distance $\Vert x_1-x_2\Vert_2$, which makes the discretized field realizations vary highly even across adjacent grid points (see Figure \ref{sec5_fig:exper3_K}). In the following two subsections, we will take $l=0.1$ and KLE are used to sample from $\kappa(x,\xi)$, i.e.,
\begin{equation}
	\xi(x) = \sum_{i=1}^Q \sqrt{\lambda_i}z_ig_i(s),
	\label{sec5_eq:KLE}
\end{equation}
where $\lambda_i$ and $g_i(x)$ are the eigenvalues and eigenfunctions of the covariance function $k(\cdot,\cdot)$, $z_i$ are i.i.d standard norm, and $Q$ is the number of expansion terms that is used to control the intrinsic dimensionality.

\subsubsection{Nonlinear source terms}
\begin{figure}[ht]
	\centering
	\includegraphics[scale=0.23]{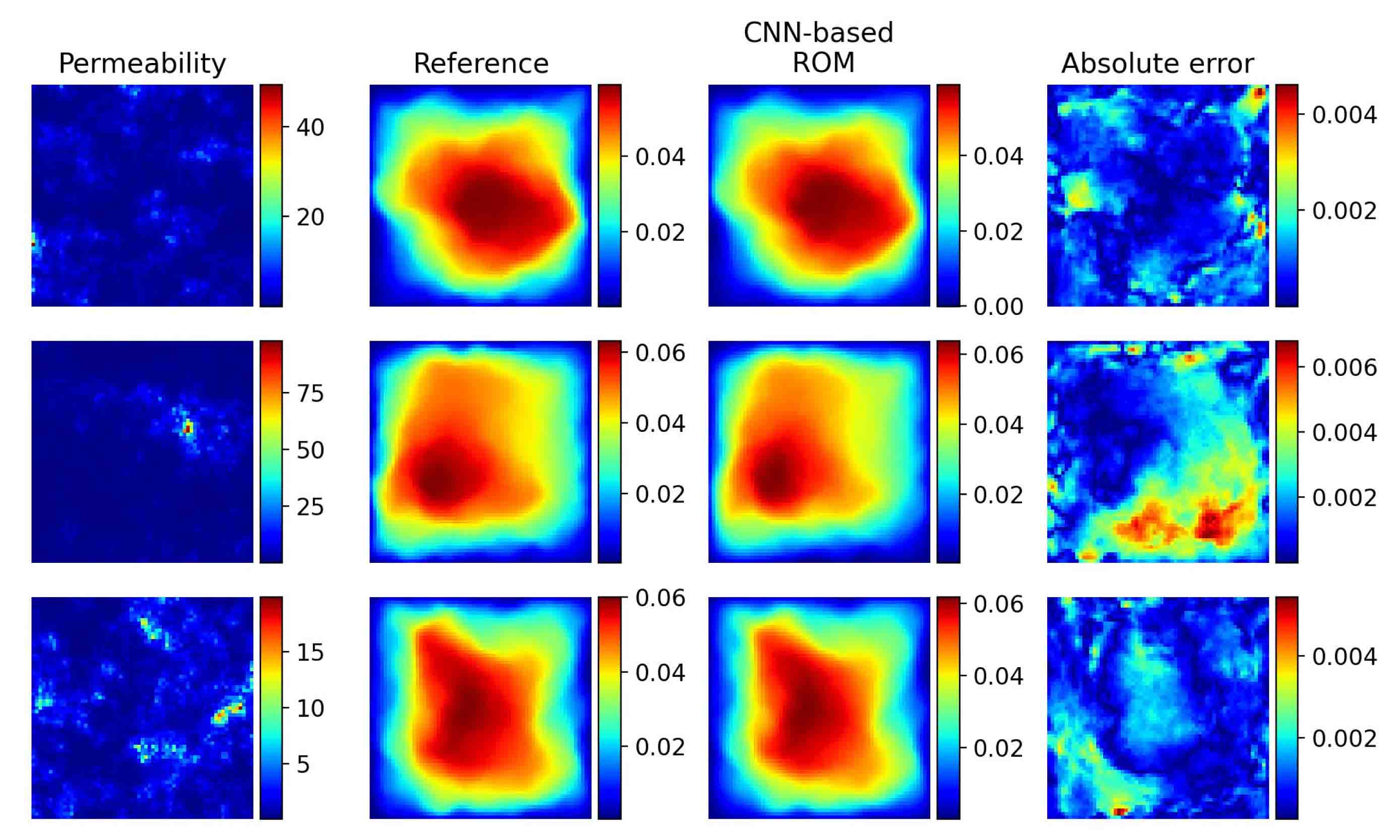}
	\caption{The predictive results of CNN-based ROM with ten basis functions at three test instances of $K$. }
	\label{sec5_fig:exper3_solution}
\end{figure}

We first consider the case that the source term is nonlinear as follows,
\begin{equation}
	\left\{
		\begin{aligned}
			&-\nabla \cdot\Big(\kappa(x,\xi)\nabla u(x)\Big)=\sin\Big(10\pi u(x)\Big) + \cos\Big(10\pi u(x)\Big), x \in \mathcal{S},\\
			&u(x)=0, x \in \partial{\mathcal{S}}.
		\end{aligned}
		\right.
	\label{sec5_eq:exper3_equation}
\end{equation}
The weak form of above equation is that find $u \in V$, such that
\begin{equation}
\int_{\mathcal{S}}\kappa\nabla u \nabla v\ dx = \int_{\mathcal{S}} \Big(\sin(10\pi u) + \cos(10\pi u)\Big) \ v\ dx,\ \forall v \in V.
\end{equation}
We solve above variational problems by Newton's iteration methods on a fixed mesh with rectangular elements and $65 \times 65$ nodes. Once we get the high-fidelity solution $u_h(K)$, we can obtain entries of stiffness matrix $A_h(K)\subset \mathbb{R}^{4225\times 4225}$ and the load vector $F_h(K)\subset \mathbb{R}^{4225\times 1}$ are
\begin{equation}
	\begin{aligned}
		a_h(i,j) &= \int_{\mathcal{S}} K \nabla \varphi_i \nabla \varphi_j \ dx,\\
		f_h(i) &= \int_{\mathcal{S}} u_h(K) \nabla \varphi_i\ dx.
	\end{aligned}
	\nonumber
\end{equation}
where $\{\varphi_i,\ldots,\varphi_{4225}\}$ is the bilinear basis functions of $V_h$. Here, we use element means to approximate the value of $u_h$ at the integral points (the same as Figure \ref{sec5_fig:grid_nn}). The Basis CNNs with architecture (Table \ref{sec5_table:architecture of Basis CNN}) are first trained using 5120 data pairs. Since $u_h(K)$ is unknown for test instances, Basis CNNs themselves can not realize online computation. We further use Coef CNNs to learn the mapping from $K$ to the coefficients $c(K)$. The specific architecture of Coef CNNs is presented in Table \ref{sec5_table:architecture of Coef CNN}
\begin{table}[H]
	\centering  
	\setlength{\abovecaptionskip}{0cm}
	\setlength{\belowcaptionskip}{0.2cm}
	\caption{The architecture of Coef CNNs}  
	\label{sec5_table:architecture of Coef CNN}  
	\scalebox{0.8}{
		\begin{tabular}{ccc}
			\toprule  
			  Layers &Resolution H $\times$ W & The number of parameters \\  
			\midrule
			& & \\[-6pt]  
			Input &63$\times$ 63 &-\\
			& & \\[-6pt]  
			Conv block (25$\times$ 25,32,1) &63$\times$ 63 &20000 \\
			& & \\[-6pt]  
			Conv block (25$\times$ 25,32,2) &32$\times$ 32 &20000 \\
			& & \\[-6pt]  
			Conv block (25$\times$ 25,64,1) &32$\times$ 32 &40000\\
			& & \\[-6pt]  
			Conv block (25$\times$ 25,64,2) &16$\times$ 16 &40000\\
			& & \\[-6pt]  
			Conv block (25$\times$ 25,128,1) &16$\times$ 16 &80000\\
			& & \\[-6pt]  
			Conv block (25$\times$ 25,128,2) &8$\times$ 8 &80000\\
			& & \\[-6pt]  
			FC(300) &300$\times$ 1 & 2457600\\
			& & \\[-6pt]  
			FC(300) &300$\times$ 1 & 90000\\
			& & \\[-6pt]  
			FC(10) &10$\times$ 1 & 3000\\
			\bottomrule
		\end{tabular}
	}
\end{table}
The same dataset is used to train Coef RNN. The relative test mean error is $4.43\%$, which is estimated by 2500 test instances. Besides, several results are shown in Figure \ref{sec5_fig:exper3_solution}. Compared to traditional ROM, which need iterative methods on the online stage, we can directly obtain the prediction of solutions using CNN-based ROM.

\subsubsection{Nonlinear differential operators}
\label{ssec5:Nonlinear Differential operators}
\begin{figure}[ht]
	\centering
	\includegraphics[scale=0.23]{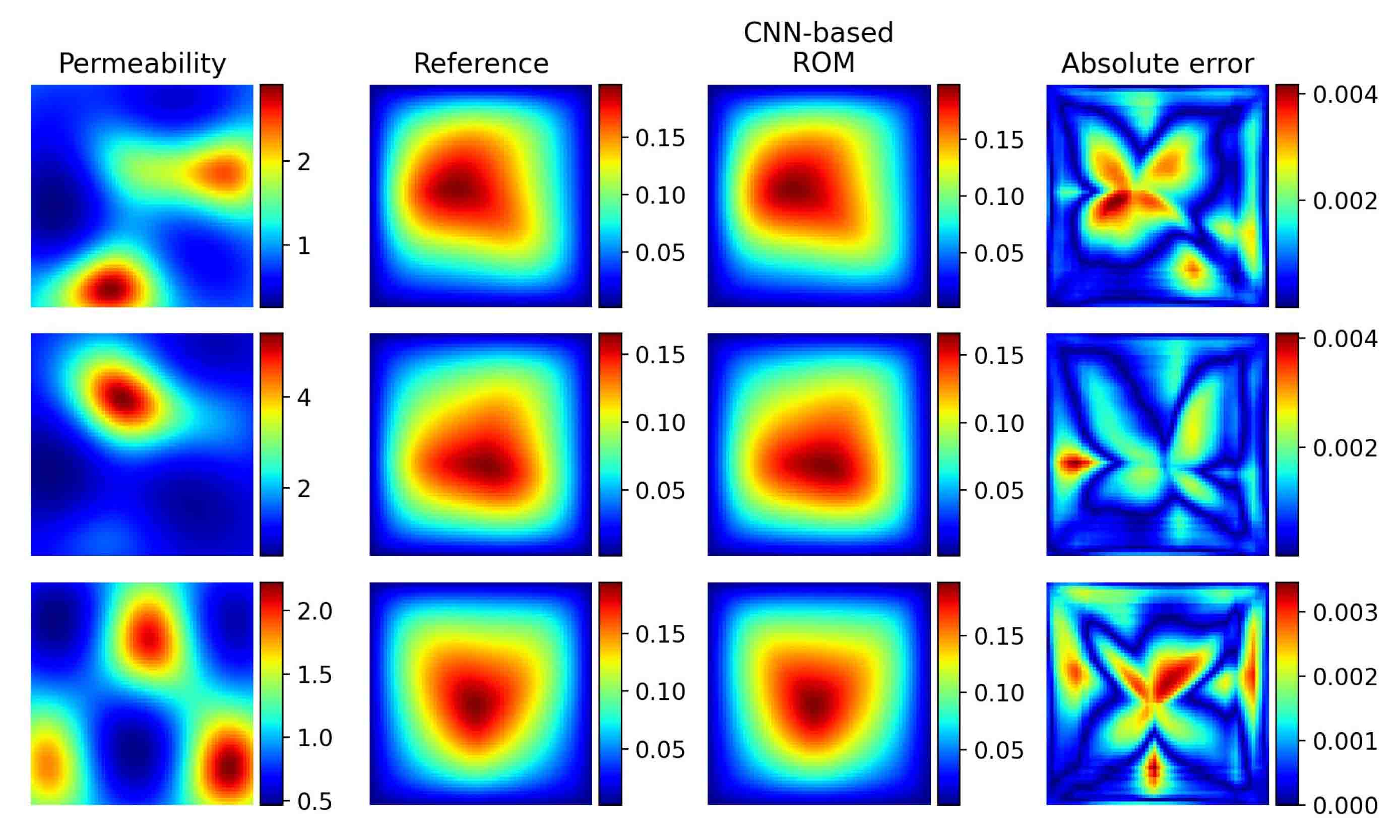}
	\caption{The predictive results of CNN-based ROM with ten basis functions at three test instances of $K$. }
	\label{sec5_fig:exper4_solution}
\end{figure}

In this section, we consider the p-Laplacian equations as follows,
\begin{equation}
		\left\{
		\begin{aligned}
			&-\nabla \cdot\Big(\kappa(x,\xi)|u(x)|^{p-2}\nabla u(x)\Big)= 1, x \in \mathcal{S},\\
			&u(x)=0, x \in \partial{\mathcal{S}}.
		\end{aligned}
		\right.
	\label{sec5_eq:exper4_equation}
\end{equation}
The p-Laplacian operator in the left hand of equation (\ref{sec5_eq:exper4_equation}) is derived from a nonlinear Darcy law and continuity equation. For p=2, equation (\ref{sec5_eq:exper4_equation}) is equivalent to the equation in Section \ref{ssec5:Application to Darcy flow}, which is linear. Here, we take $p=3$, and minimize the total potential energy of the flow, i.e.,
\begin{equation}
	u = \mathop{\arg\min}\limits_{v\in V} \frac{1}{p}\int_{\mathcal{S}}\kappa\Vert\nabla  v\Vert_2^p-\int_{\mathcal{S}}v dx.
	\nonumber
\end{equation}
We approximate $u$ on the mesh with triangular elements and $65\times 65$ nodes. The above optimization problems are solved for 12740 samples of the log-Gaussian random field with bandwidth 0.1 by an efficient algorithm \cite{sec5:plaplace}, among which 10240 instances are split to training dataset and the rest are used for test. The stiffness matrix $A_h(K)\in \mathbb{R}^{4225\times 4225}$ is computed by assuming that $u(K)$ is accurately approximated by $u_h(K)$, which has entries
\begin{equation}
	a_h(i,j)=\int_{\mathcal{S}}K|u_h(K)|^{p-2}\nabla\varphi_i\nabla\varphi_j \ dx,
	\nonumber
\end{equation}
where $\{\varphi_i,\ldots,\varphi_{4225}\}$ is the piecewise linear basis functions of $V_h$. The definition of load vector $F_h(K)\in \mathbb{R}^{4225\times 1}$ is the same as equation (\ref{sec5_eq:exper1_fi}). We train both Basis CNNs and Coef CNNs by Adam for 50 epochs. The relative test mean error is $1.49\%$ and several test instances are shown in Figure \ref{sec5_fig:exper4_solution}.

\subsubsection{Application to inverse problems}
\begin{figure}[H]
	\centering
	\includegraphics[scale=0.12]{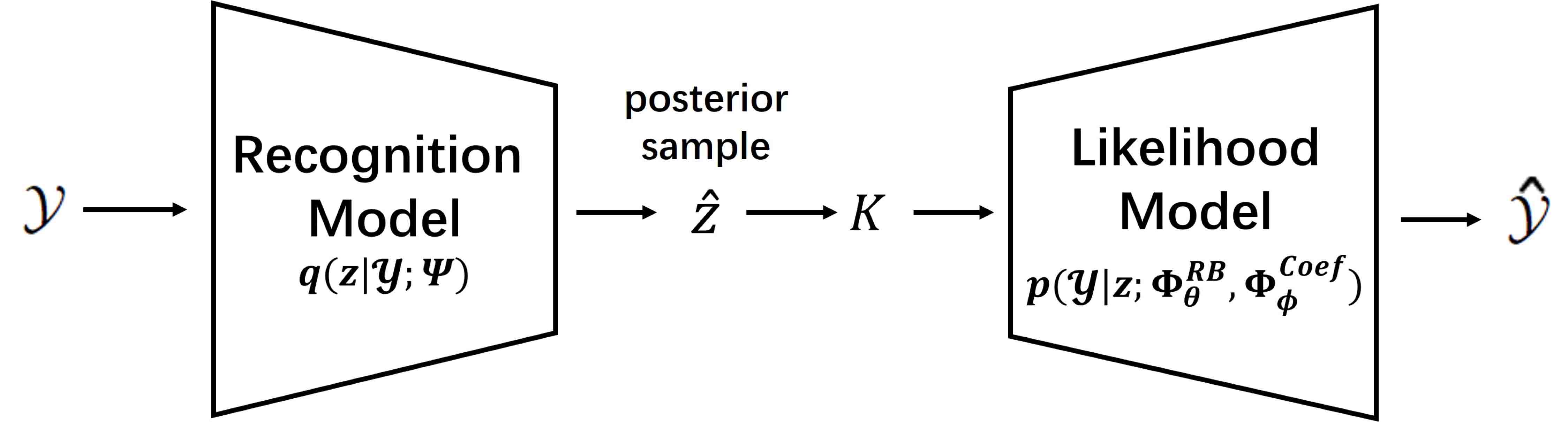}
	\caption{The schema of Surrogate-constrained VAE.}
	\label{sec5_fig:exper4_VAE}
\end{figure}
As shown in above numerical results, the proposed method provides an efficient and precise surrogate model for the random multiscale equation (\ref{sec2_eq:problem setup}). In this section, we will apply CNN-based ROM to inverse problems. Given dataset $\mathcal{Y}=\{y_1,y_2,\ldots,y_{N_{obs}}\}$ observed by $N_{obs}$ sensors on the spatial points $\{x_1,x_2,\ldots,x_{N_{obs}}\}$, we want to estimate the coefficient vector $\boldsymbol{z}:=[z_1,\ldots,z_Q]$ of KLE expansion of the Gaussian random fields defined in equation (\ref{sec5_eq:KLE}).

A method that combines Variational Autoencoder (VAE) \cite{sec5:VAE} with CNN-based ROM is introduced to solve this problem. The schema of this method is presented in Figure \ref{sec5_fig:exper4_VAE}. VAE is a type of generative model that belongs to the family of Autoencoders, specifically designed for learning latent representation of the input data and generate new data samples. It is composed of the recognition model, which learn the mapping from input data to parameters of posteriors for latent variables, and the likelihood model, which reconstructs data from latent space. In standard VAE, both recognition model and likelihood model are represented by neural networks. As an unsupervised method, VAE learn the latent variables purely by data, which is not suitable for inverse problems. Hence, we use the trained CNN-based ROM as parts of likelihood model to constrain the learning of latent variables, which is inspired by PDE-constrained Bayesian inversion methods \cite{sec5:Bayesian inverse problem}. Therefore, we call the method Surrogate-constrained VAE and the unknown parameters $\Psi$ are only contained in the recognition model, that is
\begin{equation}
	\psi = \Phi_{\Psi}^{Recog}(\mathcal{Y}),
	\nonumber
\end{equation}
where $\psi$ are the local variational parameters that define the posterior distributions $q(\boldsymbol{z}|\mathcal{Y};\psi)$ that approximate the true posterior $p(\boldsymbol{z}|\mathcal{Y})$. The mapping $\Phi_{\Psi}^{Recog}$ is a fully-connected neural networks. The problem thus becomes solving for the global variational parameters $\Psi$. Using such a neural network, we can obtain the posterior estimation of coefficients $z_i$ by passing the observation data $\mathcal{Y}$ through $\Phi_{\Psi}^{Recog}$, which can not be achieved by PDE-constrained Bayesian inversion methods.

\begin{figure}[ht]
	\centering
	\includegraphics[scale=0.14]{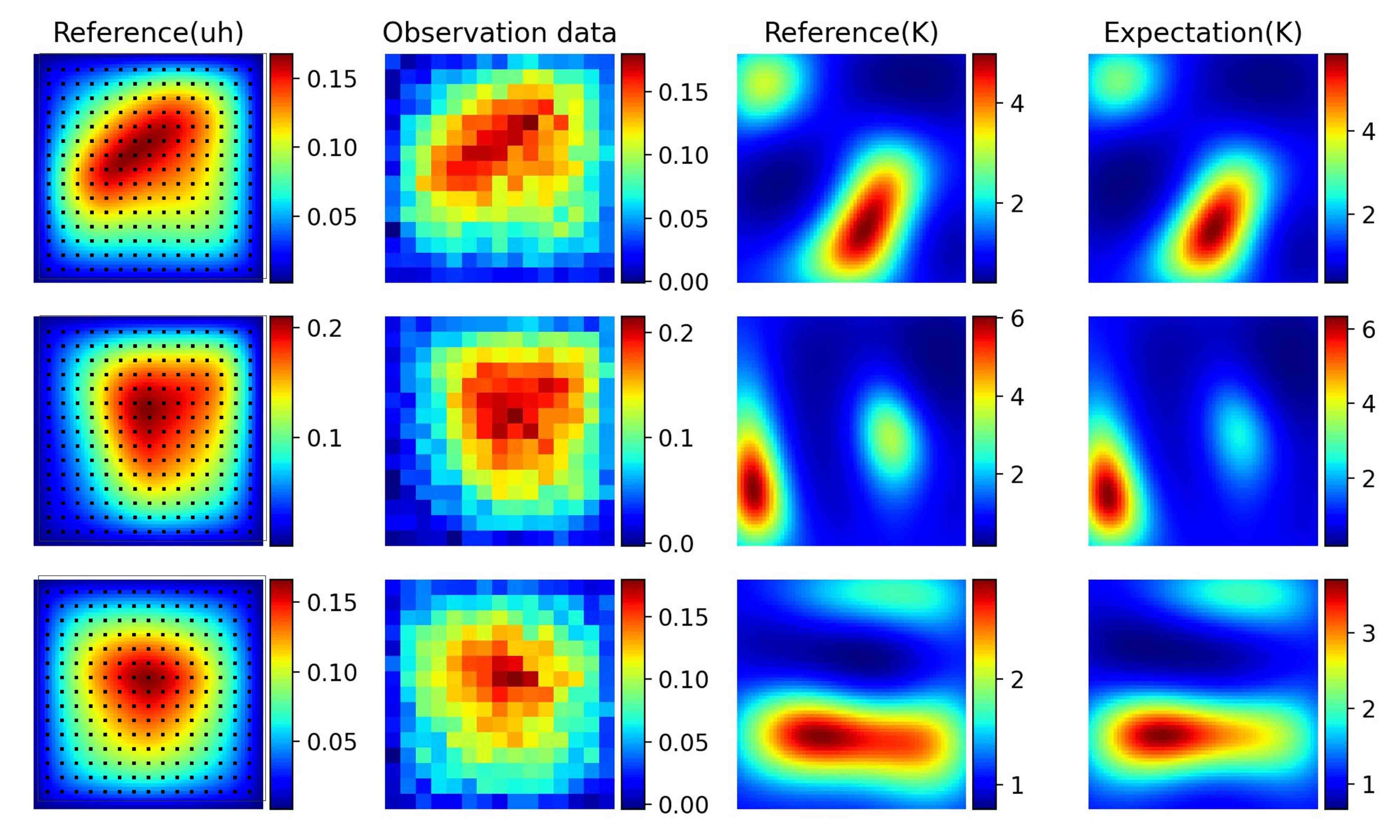}
	\caption{The expectations of random inputs, which is estimated by surrogate-constrained VAE with $15\times 15$ sensors ($\sigma_{obs}=0.01$).}
	\label{sec5_fig:exper4_K}
\end{figure}

\begin{figure}[H]
	\centering
	\includegraphics[scale=0.23]{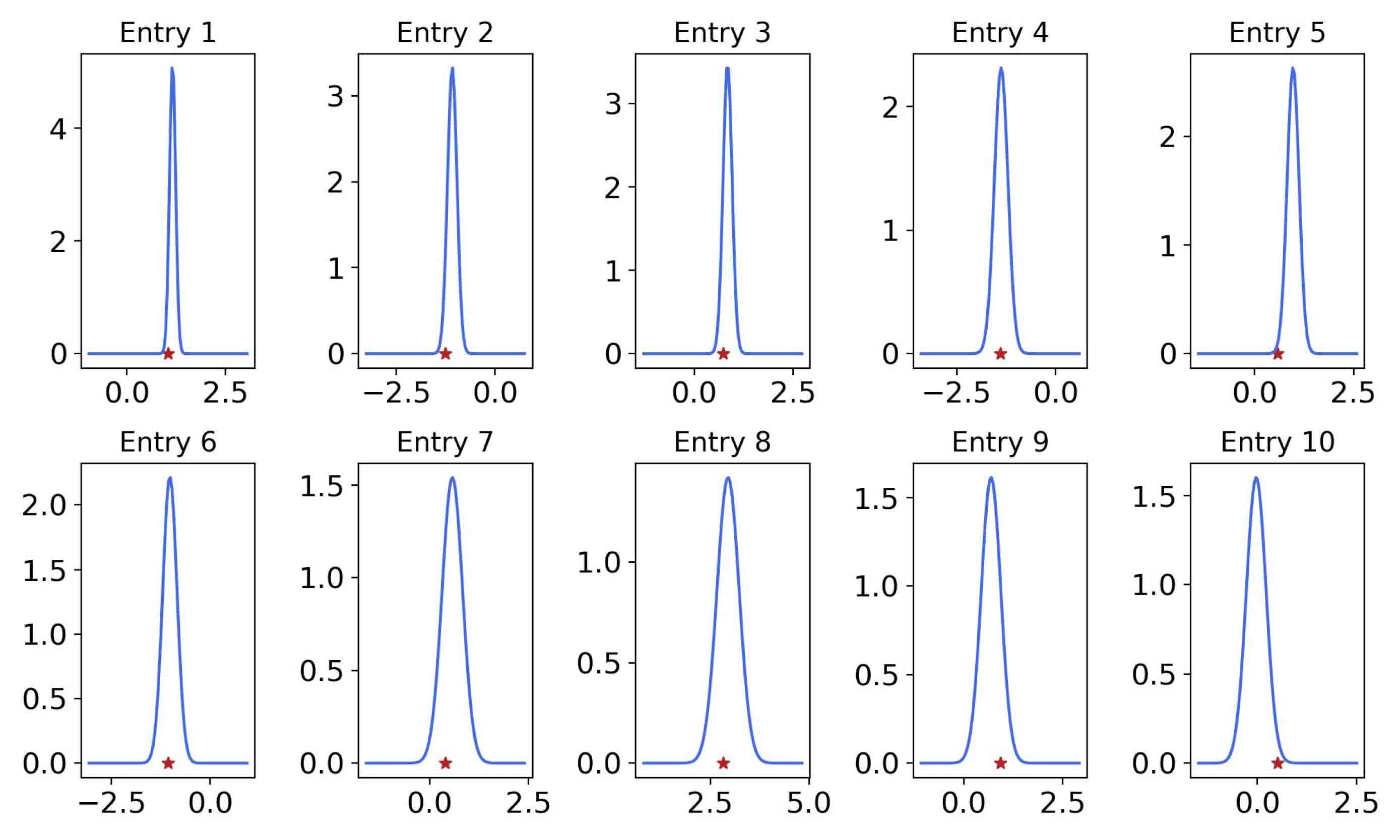}
	\caption{The posterior density of coefficients $z_i$. The number of sensors is $225$ and the observation noise is at the level of 0.01. The true coefficients are denoted by red stars, and the posterior densities of the estimated $z_i$ are represented by blue lines.}
	\label{sec5_fig:exper4_pdf}
\end{figure}

\begin{figure}[H]
	\centering
	\includegraphics[scale=0.14]{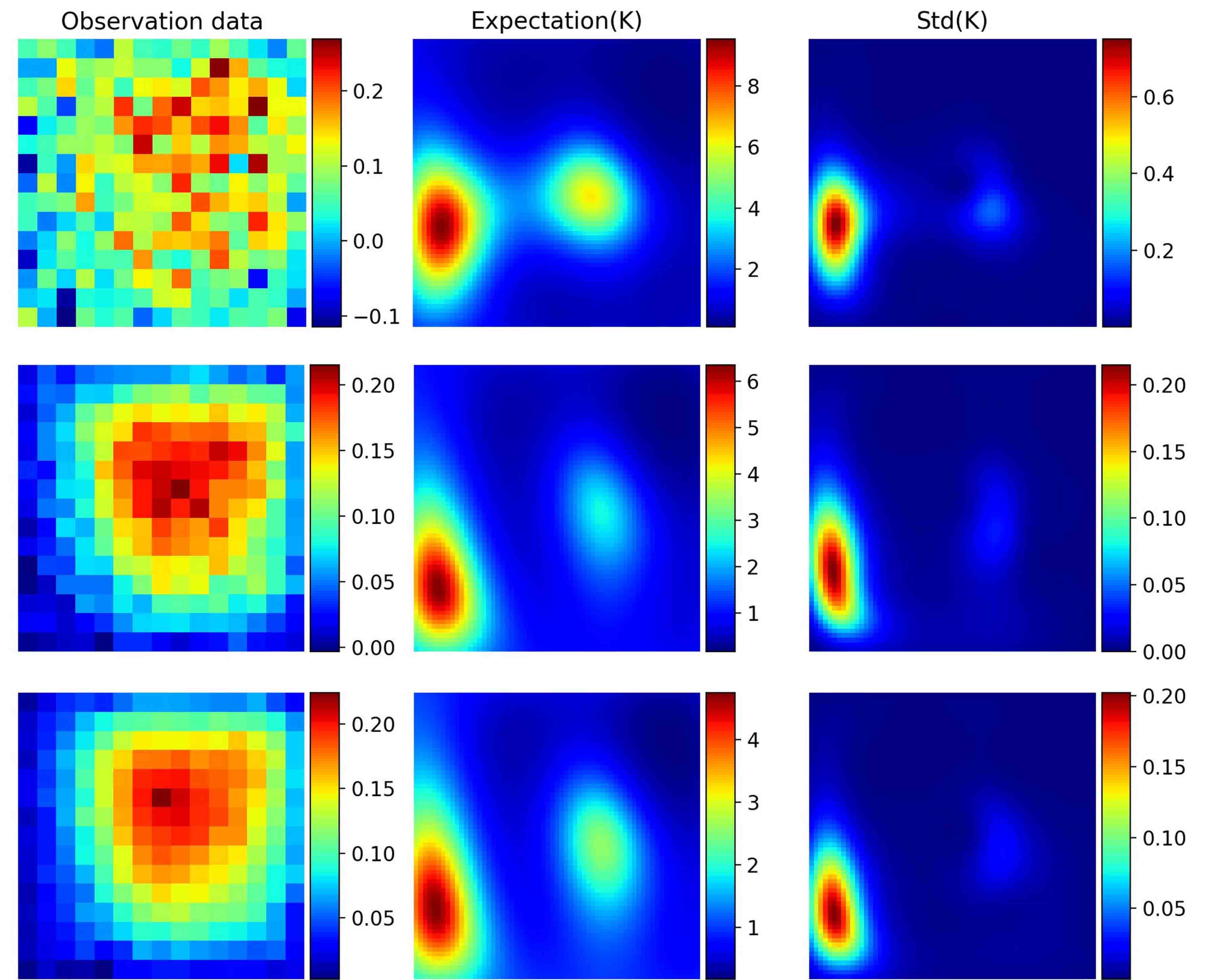}
	\caption{The expectations and standard deviations of random inputs, which is estimated by surrogate-constrained VAE with $8\times 8$ sensors. The observation noises from top line to bottom line are at the level of 0.05, 0.01, 0.005, respectively.}
	\label{sec5_fig:exper4_different_noise}
\end{figure}

Next, we will use variational inference to learn the parameters $\Psi$. Because the likelihood of data $p(y)$ is hard to computed, we instead minimize the lower bound of it that is known as evidence lower bound (ELBO),
\begin{equation}
	\text{ELBO}(\Psi) = \mathbb{E}_{q(\boldsymbol{z}|\mathcal{Y};\psi)}\Big(\log\ p(y|\boldsymbol{z};\Phi_{\theta}^{Base},\Phi_{\phi}^{Coef},\sigma_{obs})\Big)+ \text{KL}\Big(q(\boldsymbol{z}|\mathcal{Y};\psi)\Vert p(\boldsymbol{z})\Big),
	\label{sec5_eq:exper4_ELBO}
\end{equation}
where $p(y|\boldsymbol{z};\Phi_{\theta}^{Base},\Phi_{\phi}^{Coef},\sigma_{obs})$ is the likelihood model with i.i.d Gaussian observation noise $\sigma_{obs}$, i.e.,
\begin{equation}
	\mathcal{Y}_i = \Phi_{\theta}^{Base}(K_i)\Phi_{\phi}^{Coef}(K_i)+\epsilon_i, \epsilon_i \sim \mathcal{N}(0,\sigma_{obs}^2 I).
	\nonumber
\end{equation}
And $p(\boldsymbol{z})$ is the prior, we take it the standard Gaussian distribution because the coefficients of KLE of Gaussian random field is distributed to i.i.d Gaussian distribution. The approximation posterior distribution $q(\boldsymbol{z}|\mathcal{Y};\psi)$ are also Gaussian with mean vector $\mu$ and covariance matrix $\sigma^2 I$, i.e.,
\begin{equation}
	\boldsymbol{z}|\mathcal{Y} \sim \mathcal{N}(\mu,\sigma^2 I).
	\nonumber
\end{equation}
Furthermore, we can understand the equation (\ref{sec5_eq:exper4_ELBO}) from the perspective of inverse problems. The first term aims at maximizing the likelihood of the reconstructed data. The second term is the Kullback-Leibler divergence of two probability density, which is defined as
\begin{equation}
	\text{KL}(p \Vert q)=\int p(x) \log\frac{p(x)}{q(x)} \ dx.
	\nonumber
\end{equation}
It works as the regularization term such that posterior is as similar as possible to prior, which can help to alleviate the ill-posedness of the inverse problem. In terms of reparametrization tricks \cite{sec5:VAE}, we can use gradient descent algorithms, such as Adam, to optimize neural network parameters $\Psi$. Once the recognition model is trained, we can also estimate the expectations of $K$ by Monte Carlo method, i.e.,
\begin{equation}
	\mathbb{E}_{q(\boldsymbol{z}|\mathcal{Y};\psi)} (K)
	\approx \frac{1}{M}\sum_{i=1}^M \exp\Big(\sum_{j=1}^Q \sqrt{\lambda_j}z_j^{(i)}\boldsymbol{g}_j\Big), \boldsymbol{z}\sim q(\boldsymbol{z}|\mathcal{Y};\psi).
	\label{sec5_eq:expr4_expectation}
\end{equation}
where $\boldsymbol{g}_j$ is the discrete vector of eigenfunctions $g_j$ in KLE (\ref{sec5_eq:KLE}). Similarly, we can also obtain the variance of $K$ based on estimations (\ref{sec5_eq:expr4_expectation})
\begin{equation}
	Var_{q(\boldsymbol{z}|\mathcal{Y};\psi)} (K)
	\approx \frac{1}{M}\sum_{i=1}^M \Bigg( \exp\Big(\sum_{j=1}^Q \sqrt{\lambda_j}z_j^{(i)}\boldsymbol{g}_j\Big) -\mathbb{E}_{q(\boldsymbol{z}|\mathcal{Y};\psi)} (K)\Bigg)^2, \boldsymbol{z}\sim q(\boldsymbol{z}|\mathcal{Y};\psi).
	\label{sec5_eq:expr4_variance}
\end{equation}
Finally, we will demonstrate the performance of Surrogate-constrained VAE for p-Laplacian equation in Section \ref{ssec5:Nonlinear Differential operators}. We first place $15\times 15$ sensors uniformly in the spatial space as shown in the first column of Figure \ref{sec5_fig:exper4_K} and assume that the observation noise is 0.01. We can see from Figure \ref{sec5_fig:exper4_K} that Surrogate-constrained VAE can accurately estimate the expectation of permeability $K$. The posterior density of coefficients $z_i$ of the test instance in the first row of Figure \ref{sec5_fig:exper4_K} are further drawn in Figure \ref{sec5_fig:exper4_pdf}. We also change the noise level, we can see from Figure \ref{sec5_fig:exper4_different_noise} that the standard deviations decrease as the noise level decrease.

\begin{figure}[ht]
	\centering
	\includegraphics[scale=0.14]{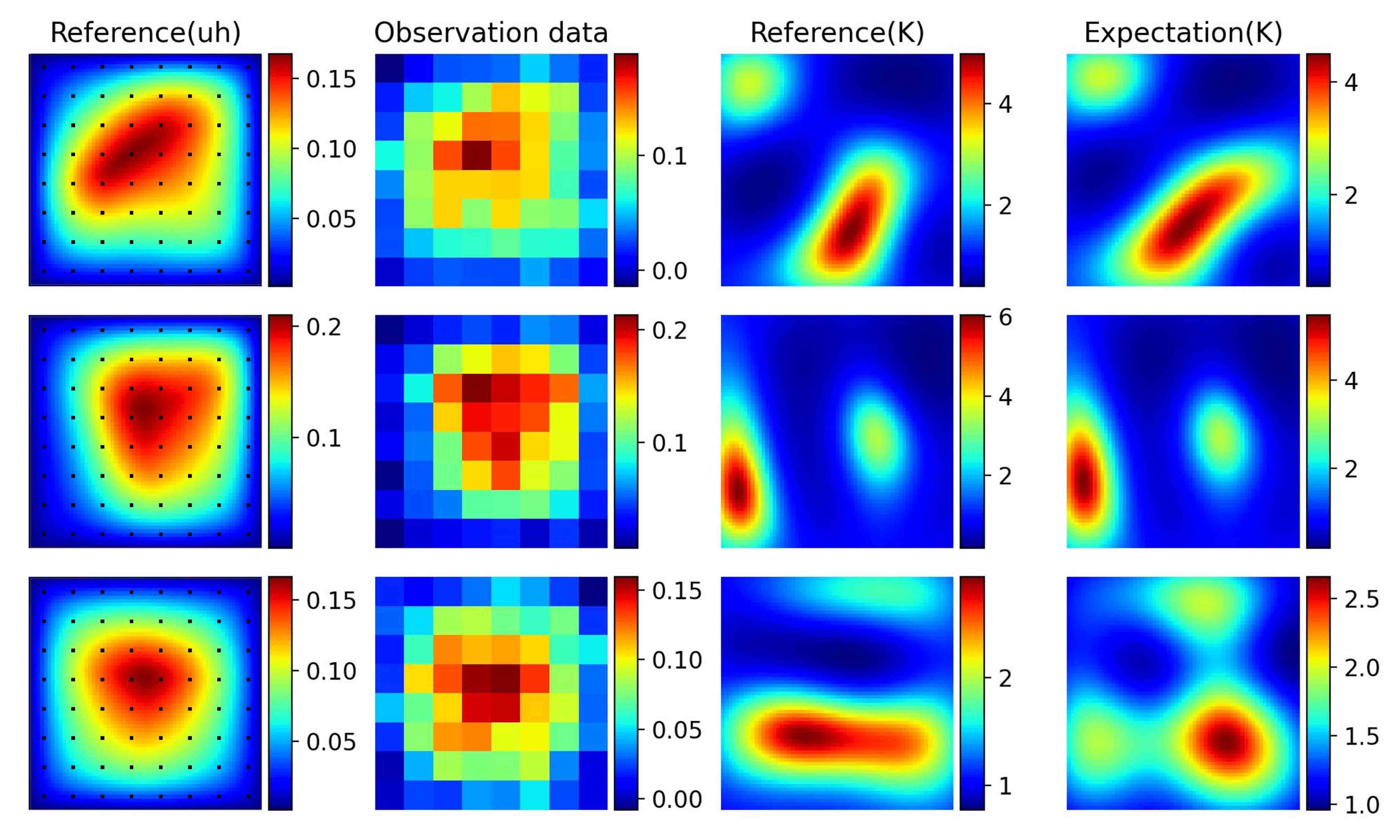}
	\caption{The expectations of random inputs, which is estimated by surrogate-constrained VAE with $8\times 8$ sensors ($\sigma_{obs}=0.01$).}
	\label{sec5_fig:exper4_K_0.01_64}
\end{figure}

\begin{figure}[H]
	\centering
	\includegraphics[scale=0.23]{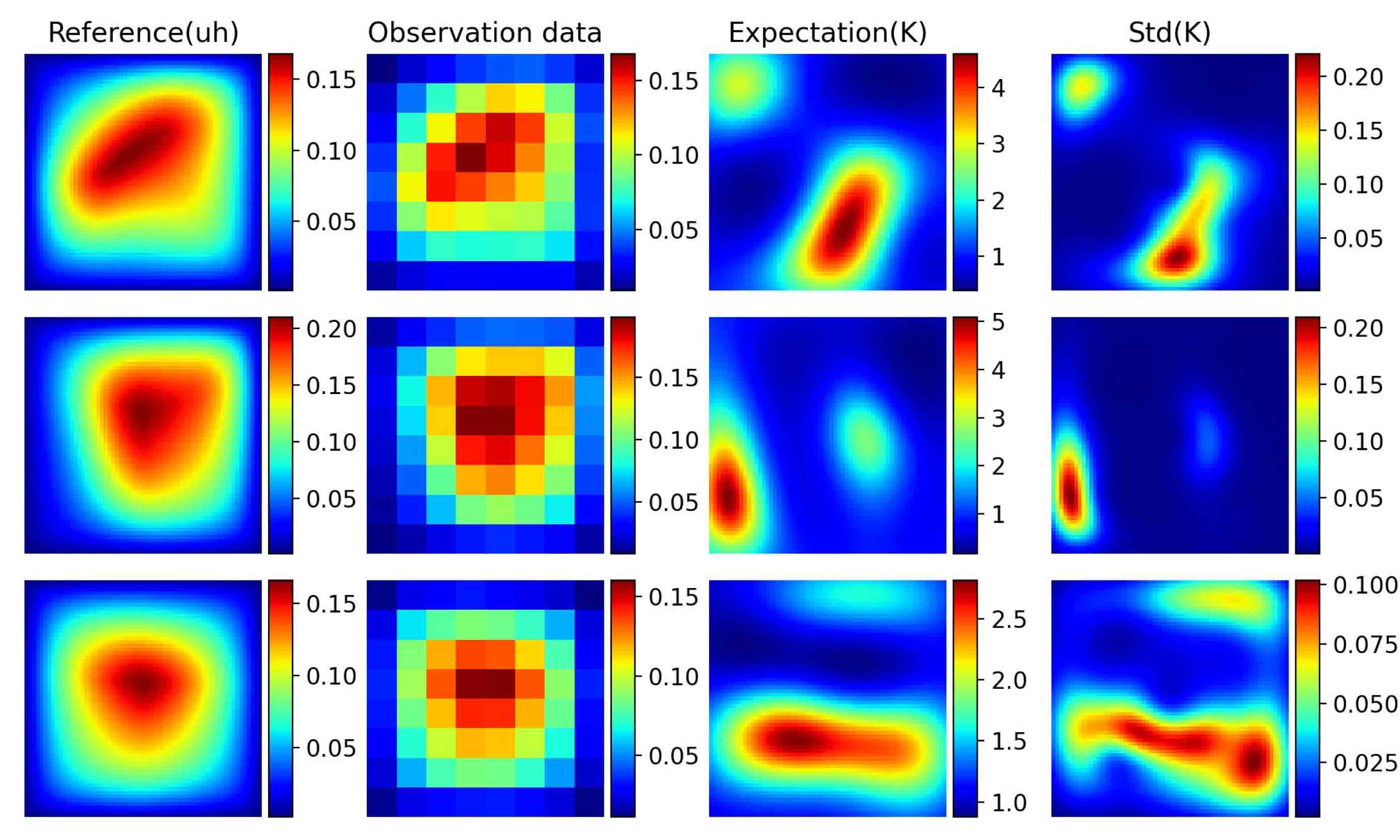}
	\caption{The expectations and standard deviations of random inputs, which is estimated by surrogate-constrained VAE with $8\times 8$ sensors ($\sigma_{obs}=0.001$).}
	\label{sec5_fig:exper4_K_0.001_64}
\end{figure}

To further illustrates performance of the proposed method, we fix the noise level  and use more sparse observation data with $8\times 8$ sensors. The estimation in Figure \ref{sec5_fig:exper4_K_0.01_64} only captures the basic shapes of permeability due to the high noise level. We thus decrease the noise level to 0.001, the performance, shown in Figure \ref{sec5_fig:exper4_K_0.001_64}, is improved as expected. It is observed that the standard deviations will be higher for less accurate estimation.

\section{Conclusion}
In this paper, we have provided a data-driven ROM methods for multiscale problems with high-dimensional random inputs, which was called CNN-based ROM. After reviewing the multiscale reduction methods and reduced basis methods, we first summarized the ROM as two main parts: the construction of reduced-order basis functions and the computation of coarse-grid, which inspired the designs of two distinct CNNs in the proposed method.

The first CNN was called Basis CNNs, which was used to learn input-specific basis functions for linear model reduction. At the output layer, activation function defined by Galerkin projection were utilized to reconstruct fine-scale solutions and constrained the learning of basis functions. On the one hand, such settings could break the limitations of the RB methods for high-dimensional problems. On the other hand, compared to multiscale reduction methods, there is no need for computing a lot of local problems online to obtain the basis functions. Because the reduced-order model learned by CNNs might be ill-conditioned, we used condition number under Frobenious norm to guarantee the stability of training. In addition, Numerical results showed that the basis functions learned by Basis CNNs were similar to fine-scale solutions, thus the number of reduced-order basis could be less than POD-based RB methods. Moreover, ROM assisted by deep learning techniques was less sensitive to data fluctuation caused by numerical methods than traditonal ROM. Fine-scale stiffness matrix and load vector were needed in Basis CNNs, which was computationally expensive and not available for nonlinear problems at the online stage. To overcome these limitations, we designed the second CNN called Coefficient CNN (Coef) to learn the coefficients of linear combination.

We also provided two applications of CNN-based ROM. We first explored the connections between the proposed method and MsFEM. It was shown that CNN-based ROM could be used for learning MsFEM basis functions within large oversampling regions. Thus, the learned basis functions could be used for different source terms. In the last subsection of numerical results, it was  demonstrated that CNN-based ROM could be used as surrogates for solving inverse problems of the p-Laplacian equation with Gaussian random field.

At the end, we would like to remark some limitations of our methods, which can be further investigated in the future. Firstly, basis functions learned by CNN-based ROM itself can not be used for different boundaries and different source terms. Secondly, the dataset consists of fine-scale solutions is not easy to obtain, ROM based on mixed data of coarse-scale and fine-scale simulations can be considered. Finally, we will extend our methods to the dynamical systems.

\end{document}